\documentclass[final,5p,times,twocolumn]{elsarticle}
\usepackage{graphicx}

\usepackage{amssymb}
\usepackage{lipsum}
\usepackage{xcolor}

\usepackage{mathtools}

\usepackage[normalem]{ulem}
\useunder{\uline}{\ul}{}
\usepackage{booktabs}

\usepackage{bbding}
\usepackage{pifont}
\usepackage{amssymb}
\usepackage{utfsym}

 \usepackage{multirow}

\usepackage{wasysym}

\usepackage{amsmath,amssymb}

\newcommand{\halfcheck}{\checkmark\kern-1.1ex\raisebox{.7ex}{\rotatebox[origin=c]{125}{--}}}

\setcitestyle{authoryear,colon,open={},close={}}

\usepackage[colorlinks]{hyperref}
\hypersetup{hidelinks}

\newcommand\citenew[1]{(\textcolor{cyan}{\citep{#1}})}

\journal{Journal of Network and Computer Applications}

\begin{document}

\begin{frontmatter}



\title{A Survey of Network Protocol Fuzzing: Model, Techniques and Directions}



            

\author[2]{Shihao Jiang}
\ead{jiangsh@std.uestc.edu.cn}

\author[1]{Yu Zhang}
\ead{fuzzingzyyy@gmail.com}

\author[1]{Junqiang Li}  
\ead{lijunqiang@std.uestc.edu.cn}

\author[1]{Hongfang Yu}  
\ead{yuhf@std.uestc.edu.cn}

\author[1]{Long Luo}  
\ead{llong@std.uestc.edu.cn}

\author[1]{Gang Sun \corref{cor1}}  
\ead{gangsun@std.uestc.edu.cn}


\address[1]{School of Information and Communication Engineering, University of Electronic Science and Technology of China, Chengdu 611731, China}
\address[2]{School of Computer Science and Engineering, University of Electronic Science and Technology of China, Chengdu 611731, China}

\cortext[cor1]{Corresponding author}

\begin{abstract}
As one of the most successful and effective software testing techniques in recent years, fuzz testing has uncovered numerous bugs and vulnerabilities in modern software, including network protocol software. In contrast to other fuzzing targets, network protocol software exhibits its distinct characteristics and challenges, introducing a plethora of research questions that need to be addressed in the design and implementation of network protocol fuzzers. While some research work has evaluated and systematized the knowledge of general fuzzing techniques at a high level, there is a lack of similar analysis and summarization for fuzzing research specific to network protocols. This paper offers a comprehensive exposition of network protocol software's fuzzing-related features and conducts a systematic review of some representative advancements in network protocol fuzzing since its inception. We summarize state-of-the-art strategies and solutions in various aspects, propose a unified protocol fuzzing process model, and introduce the techniques involved in each stage of the model. At the same time, this paper also summarizes the promising research directions in the landscape of protocol fuzzing to foster exploration within the community for more efficient and intelligent modern network protocol fuzzing techniques.
\end{abstract}


\end{frontmatter}


\section{Introduction}
\label{introduction}

Network protocol is a set of specifications that governs communication between devices over a computer network. It defines the format, sequence, and error handling of messages exchanged on the network and is critical to enabling mutual communication between devices in network infrastructures \citenew{hermann1995framework}. Based on these standardized specifications, network protocol software is implemented, which can complete complex functions such as connection establishment, data analysis, and error detection during computer network communication and data exchange. During execution, the network protocol software will expose its network interface and handle normal, malformed, or even malicious communication traffic according to the rules specified by the protocol \citenew{4}. Unfortunately, due to the diversity of requirements, network protocol software is often complex and difficult to implement completely correctly. Moreover, a significant portion of network protocol software is implemented based on C/C++, which is known to be memory-unsafe. Consequently, widespread vulnerabilities have posed serious security issues to the network protocol software stack. For instance, the infamous Heartbleed \citenew{heartbleed} security vulnerability affecting thousands of servers around the world is caused by the incorrect handling of length parameter boundary checks in OpenSSL (open source software that implements the SSL protocol). Research indicates that the prevalence of security vulnerabilities in network protocol software continues to exhibit rapid proliferation \citenew{stateafl}, thereby intensifying the demand within the security community for advanced techniques in the efficient analysis of protocol software security.

\begin{table*}[ht]
\label{table1}
\centering
\resizebox{\linewidth}{!}{

\begin{tabular}{ccccccc}
\multicolumn{7}{l}{\textbf{Table 1}}     \\

\multicolumn{7}{p{20cm}}{List of contributions by survey papers. A \usym{1F5F8} denotes that the paper has a detailed summary or discussion on the topic, and a  \halfcheck denotes that the paper has a high-level discussion of the topic, but ignores the characteristics of the network protocol to varying degrees.}  \\

\specialrule{0.10em}{3pt}{3pt} 

Papers & Fuzzing fundamentals & \begin{tabular}[c]{@{}c@{}}Protocol fuzzing\\ terminology\end{tabular} & Fuzzer categorization & \begin{tabular}[c]{@{}c@{}}Unique challenges\\ in protocol fuzzing\end{tabular} & \begin{tabular}[c]{@{}c@{}}Protocol fuzzing\\ process model\end{tabular} & \begin{tabular}[c]{@{}c@{}}Review of\\ key fuzzers\end{tabular} \\ 

\specialrule{0.10em}{3pt}{3pt}

\citenew{59}  &         \usym{1F5F8}            &              \halfcheck                                                          &            \usym{1F5F8}            &                    \usym{2717}                                                             &      \halfcheck                                                                    &          \usym{1F5F8}                                                       \\
\citenew{58}             &            \usym{1F5F8}            &            \halfcheck                                                            &     \usym{1F5F8}                  &          \usym{2717}                                                                       &         \halfcheck                                                                 &                       \usym{1F5F8}                                           \\
\citenew{1}             &        \usym{1F5F8}             &                 \halfcheck                                                        &                 \usym{2717}      &     \halfcheck                                                                            &            \halfcheck                                                              &    \usym{1F5F8}                                                             \\
\citenew{munea2016network}             &     \halfcheck                 &        \usym{2717}                                                                &          \halfcheck             &                      \usym{2717}                                                            &     \usym{2717}                                                                      &            \halfcheck
\\
\citenew{daniele2023fuzzers}             &        \usym{1F5F8}              &                                                  \halfcheck                      &           \usym{1F5F8}            &                                 \halfcheck                                                &                                                \halfcheck                          &                     \usym{1F5F8}                                            \\ 
This survey             &         \usym{1F5F8}             &    \usym{1F5F8}                                                                     &      \usym{1F5F8}                  &        \usym{1F5F8}                                                                          &           \usym{1F5F8}                                                                &        \usym{1F5F8}                                                          \\
\specialrule{0.10em}{3pt}{3pt}
\end{tabular}
}
\end{table*}


Fuzz testing (Fuzzing) is one of the most efficient techniques for software security analysis, gaining widespread recognition in both academia and the industry (\textcolor{cyan}{\citep{5}};  \textcolor{cyan}{\citep{1}}). Presently, Fuzzing has evolved to proficiently detect real-world software vulnerabilities among a diverse array of targets, including file processing software (\textcolor{cyan}{\citep{aifore}};  \textcolor{cyan}{\citep{profuzzer}}), network protocol software (\textcolor{cyan}{\citep{4}};  \textcolor{cyan}{\citep{18}}), Internet of Things (IoT) firmware (\textcolor{cyan}{\citep{iotfuzzer}};  \textcolor{cyan}{\citep{snipuzz}}), operating system kernels (\textcolor{cyan}{\citep{kim2020hfl}; \textcolor{cyan}{\citep{corina2017difuze}}};  \textcolor{cyan}{\citep{bulekov2023no}}), deep learning frameworks (\textcolor{cyan}{\citep{wei2022free}};  \textcolor{cyan}{\citep{freefuzz}}), libraries (\textcolor{cyan}{\citep{zhang2021apicraft}}; \textcolor{cyan}{\citep{babic2019fudge}}), and hypervisors (\textcolor{cyan}{\citep{liu2023videzzo}}; \textcolor{cyan}{\citep{pan2021v}}). 
Fuzzing's core concept is generating a large number of test cases as input to the program under test (PUT) through automatic or semi-automatic means, all the while employing monitors to scrutinize the PUT's state for anomalies. Essentially, fuzzing aspires to automatically explore those particular inputs within the vast input space of the PUT that could potentially trigger vulnerabilities \citenew{6}. As fuzzing techniques advance, an increasing number of researchers are incorporating the characteristics of the PUT into the design of intelligent and efficient fuzzing strategies (e.g., coverage-based greybox fuzzing), aiming to alleviate the considerable time and computational resource costs associated with entirely random testing.

Since the inception of fuzzing, network protocols have consistently remained a mainstream testing target. Distinguished from other PUTs, network protocols possess unique characteristics, such as statefulness and highly structured input \citenew{hess1992unix}. These characteristics pose unique challenges to fuzzing. Traditional fuzzing techniques do not identify states and typically remain at a shallow level within the stateful protocol software, unable to find deep stateful vulnerabilities \citenew{chen2018systematic}. Furthermore, due to the highly structured nature of network protocol inputs, traditional test case generation methods can disrupt their structural integrity, resulting in a multitude of invalid inputs and reduced testing efficiency (further details will be discussed in Section \textcolor{cyan}{\ref{section2A}}). To tackle the aforementioned challenges, researchers have introduced various advanced protocol fuzzers. Unfortunately, while there are currently many papers reviewing and summarizing fuzzing techniques, few research specifically dedicated to systematizing the domain of network protocol fuzz testing. Furthermore, prior work has summarized general process models or algorithm-based descriptions of fuzzing, but these highly abstract process models overlook the uniqueness of network protocols as testing targets. We believe it is imperative to systematically review and consolidate the most cutting-edge network protocol fuzzing techniques.

In this paper, we comprehensively review high-quality literature and significant fuzzers in the field of network protocol fuzzing. We delve into the primary challenges that fuzzing encounters when applied to network protocol software and propose a generic process model in Section \textcolor{cyan}{\ref{overview}}. Then we follow the four stages in the model: protocol syntax acquisition and modeling (Section \textcolor{cyan}{\ref{syntax}}), test case generation (Section \textcolor{cyan}{\ref{testcase}}), test execution and monitoring (Section \textcolor{cyan}{\ref{execution}}), and feedback information acquisition and utilization (Section \textcolor{cyan}{\ref{Feedback Information Acquisition and Utilization}}) to analyze as well as summarize the focal points of state-of-the-art research efforts. Furthermore, we put forth promising future research directions to stimulate the continued advancement of protocol fuzzing (Section \textcolor{cyan}{\ref{direction}}). 



\paragraph{\indent\setlength{\parindent}{1em} \textbf{Related Works}}
There are many survey and review papers on fuzzing as this field of research flourishes. However, to the best of our knowledge, there is a notable scarcity of systematic reviews specifically dedicated to network protocol software fuzzing techniques within the scope of our literature selection. \citenew{59} reviews a range of state-of-the-art fuzzing techniques, summarized the process of general fuzz testing, and further investigated and classified several widely used fuzzing tools. \citenew{58} aims to unify the fuzzing field from a high-level perspective. They introduce a series of rigorously defined fuzzing taxonomies and comprehensively describe the process of general fuzzing using algorithms. \citenew{1} systematically analyzes the three gaps faced by modern fuzzing techniques from a unique perspective. It consolidates research progress in the field of fuzz testing into solutions addressing these three gaps: reducing the input space, building fuzzing theory, and achieving automatic execution. \citenew{munea2016network} initial attempt at surveying the field of protocol fuzzing, providing a summary of fuzzing terminology, and reviewing representative protocol fuzzers. However, it lacks a comprehensive overview and modeling of the protocol fuzzing process. Additionally, it lacks specificity in summarizing taxonomies related to protocol testing scenarios. \citenew{daniele2023fuzzers} provides a detailed classification and separate summaries of fuzzers targeting stateful systems. It covers a wide range of state-of-the-art fuzzers from recent years. \textcolor{cyan}{Table \ref{table1}} summarizes the contributions of these existing works.
\paragraph{\indent\setlength{\parindent}{1em} \textbf{Contributions}}
The absence of focus on network protocol fuzzing in these surveys has motivated us to engage in a comprehensive and in-depth discussion and summary in this paper. In summary, the contributions of our research paper are as follows:
\begin{itemize}
\setlength{\itemsep}{1pt}
\setlength{\parsep}{1pt}
\setlength{\parskip}{1pt}
    \item We analyze four unique challenges posed by network protocol software for the implementation of fuzzing. It's noteworthy that these challenges are highly specific to the characteristics of the protocol software itself, differing from the general challenges discussed at a higher level in existing surveys. We combine mathematical definitions with examples of protocol software to describe the characteristics of protocols and their corresponding challenges.
    \item We observe distinctions in the design and fuzzing strategies between fuzzers targeting protocol software and general fuzzers. We propose a new classification criterion for protocol fuzzers and categorize representative protocol fuzzers based on our criteria.
    \item Furthermore, we present a generic process model for protocol fuzz testing. In contrast to the unified fuzzing process summarized in existing work, our process model fully reflects the unique criteria for designing fuzzers for protocol software and distills numerous solutions proposed in recent years to address protocol-specific challenges. Moreover, we delve into each execution stage of the process model, summarizing the latest fuzzing strategies tailored for specific stages while providing both a high-level process model architecture and low-level process model details.
    \item Lastly, we discuss future promising research directions in the hope of inspiring new advancements in the field of protocol fuzzing.
\end{itemize}
\section{Overview}
\label{overview}

The foundational concept of fuzzing can be traced back to 1990 when Miller et al. subjected Unix programs to random input testing. Approximately 24\% tested programs exhibited anomalous crashes when feeding a stream of randomly generated characters, leading to the terminology ``fuzz'' for this method. The fundamental idea behind fuzzing is to employ an extensive array of random ``fuzzy inputs'' to encompass the input space of the target program, thereby identifying defects and vulnerabilities \citenew{8}.

This chapter introduces in detail the particularities and challenges of network protocol fuzzing compared with traditional software under test.  Additionally, we abstract the network protocol fuzzing process, presenting a unified process model encompassing four phases: protocol syntax acquisition and modeling, test case generation, test execution and monitoring, and feedback information acquisition and utilization.  Lastly, we introduce the classification of traditional fuzzers, proposing a categorization method more tailored to protocol fuzzers.

\subsection{Distinctive attributes and challenges inherent in network protocol fuzzing}
\label{section2A}

Unlike other fuzzing targets, network protocol software has its unique characteristics, leading to conventional fuzzers ineffective or inefficient when applied to protocol software \citenew{56}. We summarize its distinctiveness as reliance on network links, statefulness, the highly structured nature of inputs, and non-uniformity.

\paragraph{\indent\setlength{\parindent}{1em} \textbf{Reliance On Network Links}}
The input for network protocol software is the network traffic, resulting in a tightly coupled relationship between the operation and implementation of protocol software and the interfaces of network functionalities. In more detail, the network traffic processed by network protocols arrives as a sequential sequence over time, with data packets constituting the smallest units within this traffic. Hence, when testing network protocols, unlike the input for traditional software (such as simple command-line parameters or file inputs), a fuzzer needs to construct a data packet and transmit it to the protocol software through a network interface. The dependency on network links necessitates that the protocol fuzzer possesses the capability to utilize standard network interfaces (such as TCP/IP or UDP/IP) for sending and receiving network packets. Most contemporary protocol fuzzers are equipped with the capability to send and receive network packets. It is noteworthy that recent works have employed advanced approaches to overcome these constraints, as elucidated in detail in section \textcolor{cyan}{\ref{section52}}.

\begin{figure}[h]
    \centering
    \includegraphics[width=0.48\textwidth]{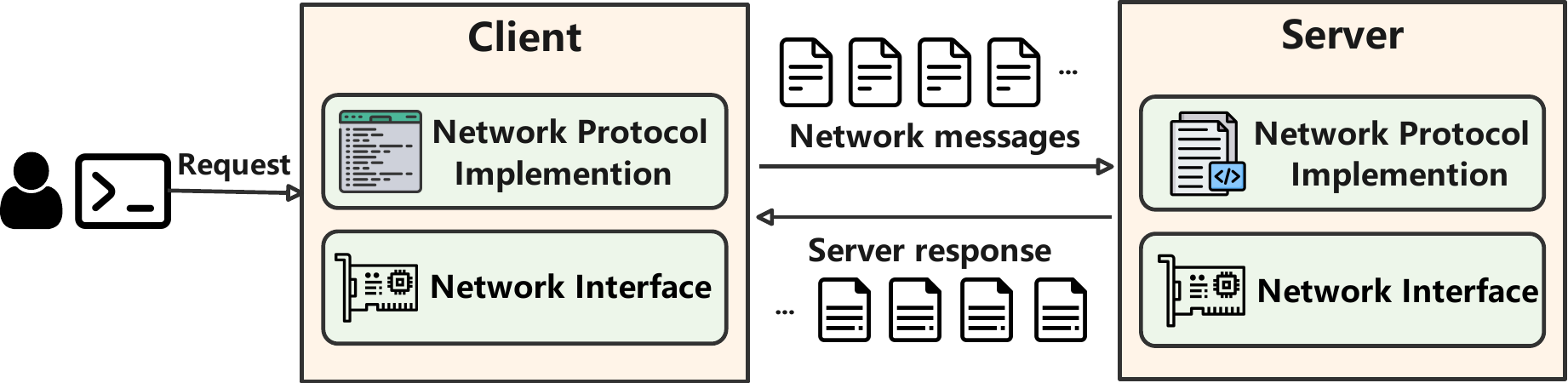}
    \caption{Protocol Software Interaction Process.}
    \label{fig:1}
\end{figure}

\begin{figure}[h]
    \centering
    \includegraphics[width=0.48\textwidth]{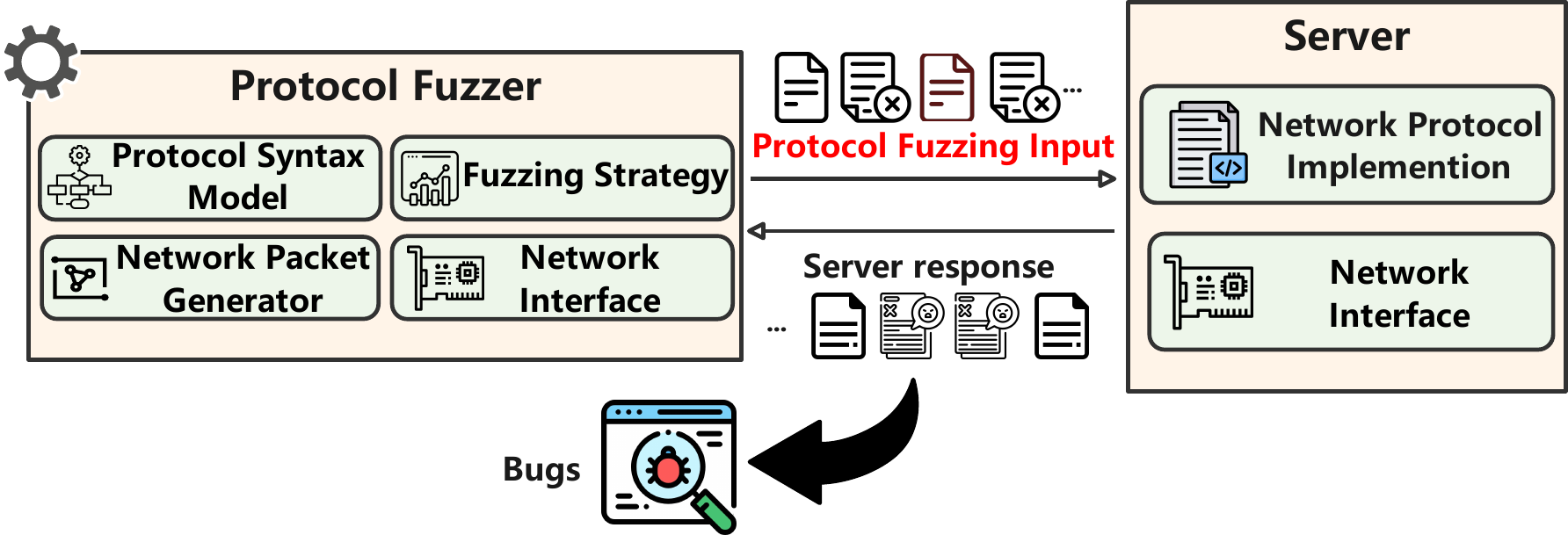}
    \caption{Protocol Fuzzing Process (for Server Side).}
    \label{fig:2}
\end{figure}

Based on these characteristics, the communication topology between the protocol fuzzer and the target closely resembles the server-client structure commonly found in network applications. During the interaction process of protocol software, the client receives user commands or receive a request from the upper level software stack, constructs corresponding data packets based on the protocol implementation, and transmits them to the server through the network link. After receiving client requests transmitted via the network interface, the server undertakes the corresponding operations and sends a response, as illustrated in Figure \textcolor{cyan}{\ref{fig:1}}. When conducting fuzzing on server-side protocol software, the fuzzer assumes the role of the client, constructing test cases through a series of components such as protocol syntax models, fuzzing strategies, and packet generators. Subsequently, the fuzzer, acting as the client, transmits the test cases to the server in the form of network packets, as depicted in Figure \textcolor{cyan}{\ref{fig:2}}. Similarly, when fuzzing client-side protocol software, the fuzzer takes on the role of the server.

\paragraph{\indent\setlength{\parindent}{1em} \textbf{Statefulness}}
Network protocol software is typically stateful. The protocol software performs corresponding operations by identifying different states during the communication process, thereby realizing complex interactions. Specifically, the state of a network protocol defines a set $ \textbf{S} = \{ I, P, O\}$, where $I$ denotes the valid input space, $O$ denotes the output space, and $P$ denotes the processing logic for inputs in the current state. Stateful protocols are in a deterministic state at any given moment throughout the communication process and change state when a specific message is received. Furthermore, the same input may result in different program behaviors when occurring in different software states. A commonly employed method for describing states and their transition logic is the finite state machine.

\begin{figure}[h]
    \centering
    \includegraphics[width=0.48\textwidth]{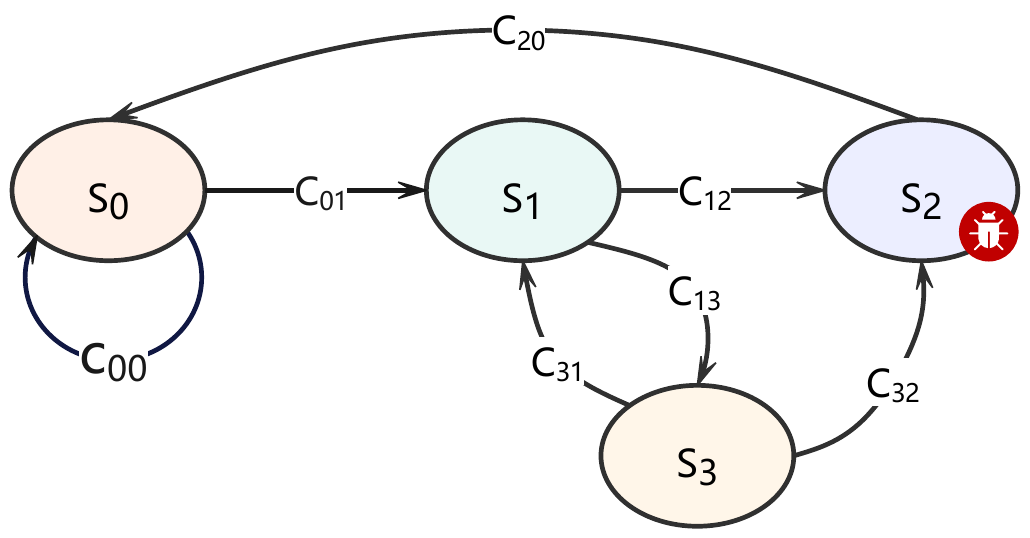}
    \caption{State Machine.}
    \label{fig:3}
\end{figure}

Taking the protocol illustrated in Figure \textcolor{cyan}{\ref{fig:3}} as an example, from a state perspective, the protocol can be abstracted as a collection of states $S$ and the associated state transition logic $T$, where $S = \{ S_0, S_1, ..., S_n \} $ and $T = \{ C_{\mathrlap{\text{01}}\phantom{0}}, C_{\mathrlap{\text{12}}\phantom{0}}, ..., C_{\mathrlap{\text{xy}}\phantom{0}} \}$. $S_i$ denotes the different protocol states, and $C_{\mathrlap{\text{xy}}\phantom{0}}$ denotes the program inputs that lead to a transition from state $S_x$ to state $S_y$. For example, the protocol program shown in the figure is initially in the $S_0$ state and enters the $S_1$ state after receiving packet $C_{\mathrlap{\text{01}}}$.

The statefulness of the protocol directly influences the design strategy of the fuzzer. It dictates that the input interacting with the protocol must be a sequence of messages with a strict chronological order (i.e., an input unit comprising multiple messages). Stateless fuzzers, in a single fuzzing iteration, provide only a single message to the protocol. This limitation confines the testing to the initial state of the protocol, resulting in highly inefficient testing.

\begin{figure}[h]
    \centering
    \includegraphics[width=0.48\textwidth]{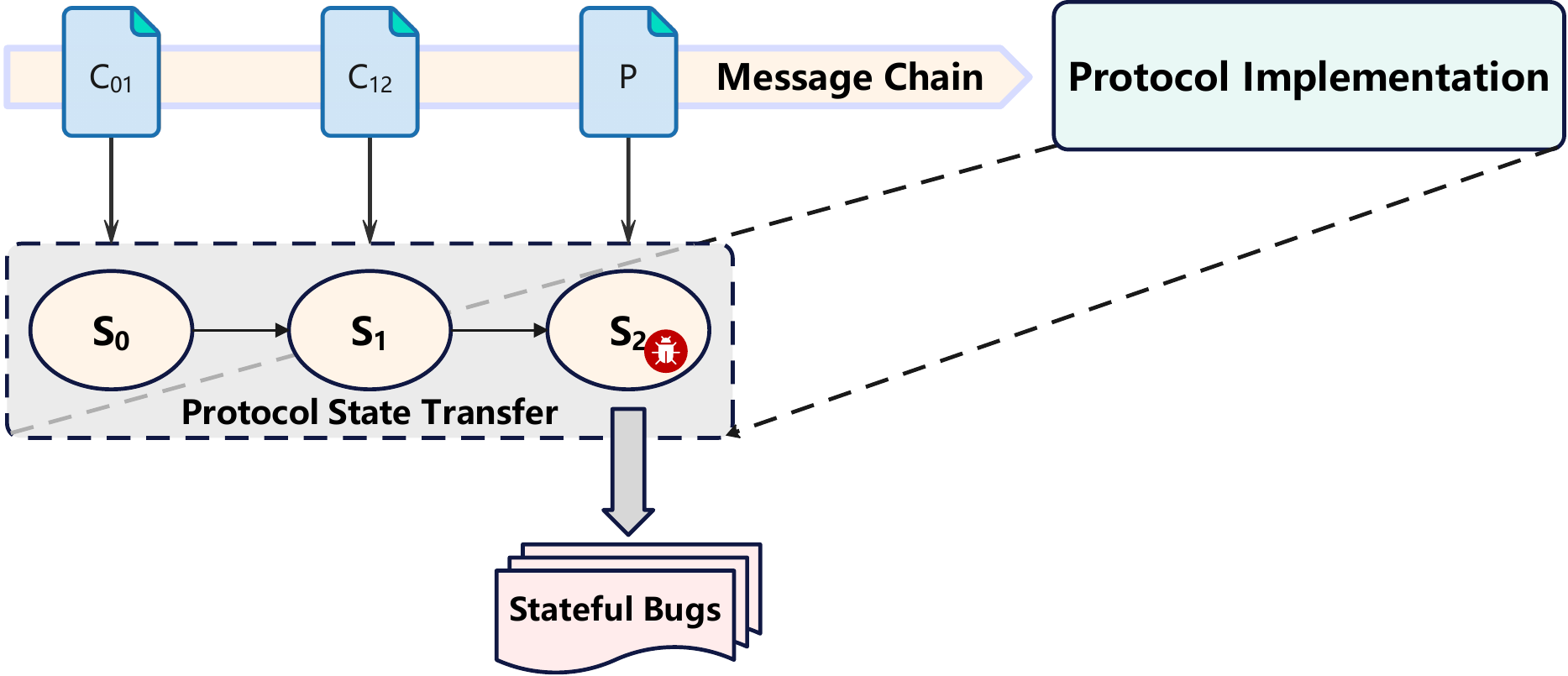}
    \caption{Stateful Bugs.}
    \label{fig:4}
\end{figure}

To fuzz the vast state space of network protocol software, fuzzers need to construct message sequences meticulously. It begins by transmitting relevant prefix messages to transition the protocol software to the target state. Subsequently, randomly crafted data packets are sent to fuzz the target state. Figure \textcolor{cyan}{\ref{fig:4}} illustrates this process, with the server reaching state $S_2$ guided by the prefix message sequence. In this state, the server receives the message P, successfully triggering a stateful bug.

\paragraph{\indent\setlength{\parindent}{1em} \textbf{Highly Structured Input}}
Network protocol messages are highly structured. Messages can be partitioned into bit or byte fields with strict grammar constraints, each with a precisely defined type and range of valid values. If fundamental grammar constraints are violated, such as the absence of the expected fixed bytes in a checksum field, the protocol software will discard the packet and terminate the connection. This outcome renders fuzzers incapable of exploring deep-seated program vulnerabilities. Therefore, fuzzers that are sensitive to input structure often have better performance in protocol fuzzing. For instance, random bit-level mutations, as employed in AFL \citenew{12}, severely disrupts the message structure of a protocol, generating a plethora of invalid test inputs \citenew{66}.

\paragraph{\indent\setlength{\parindent}{1em} \textbf{Non-uniformity}}
There are various types of network protocols, and different protocols lack uniformity in message syntax and state machines. This non-uniformity constrains the generality of protocol fuzzers, as distinct protocols may have entirely different message syntaxes and internal state machines. Existing solutions typically require additional processing to ensure coverage of protocols that can be fuzzed. For instance, the popular protocol fuzzer Peach \citenew{15} establishes generality through labor-intensive efforts, relying on manually extracted protocol specifications. AFLNet \citenew{18} is capable of fuzzing general protocol programs, requiring users to manually add codes, extract response codes from the protocol's reply messages, and write corresponding response message processing scripts.

\begin{figure*}[h]
    \centering
    \includegraphics[width=0.98\textwidth]{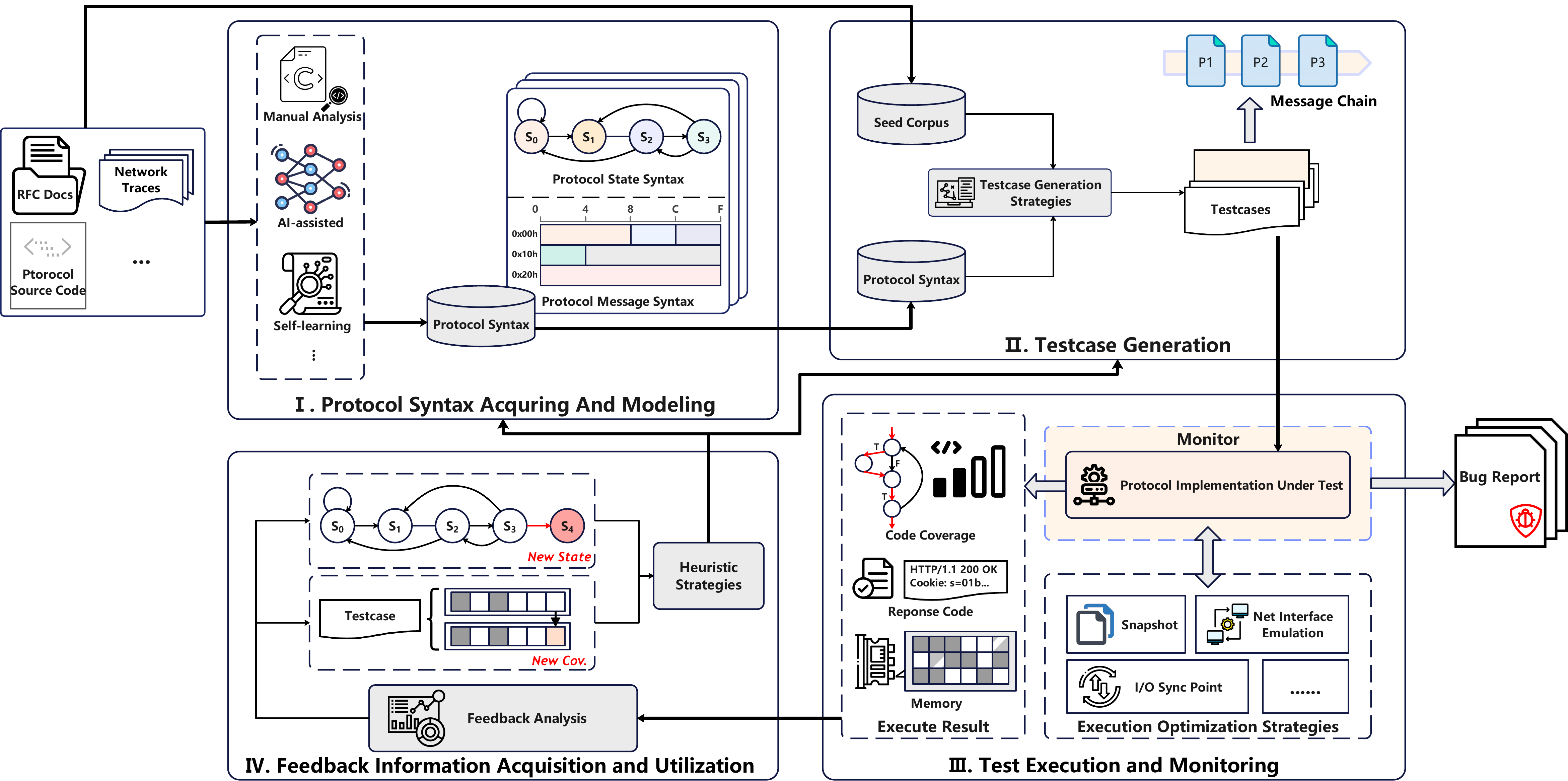}
    \caption{Unified Process Model for Network Protocol Fuzzing.}
    \label{fig:5}
\end{figure*}

\subsection{Unified process model for network protocol fuzzing}
Faced with the aforementioned challenges, researchers in the security community have proposed a series of fuzzing methods. This section outlines our modeling of a unified process model for various network protocol fuzzing approaches. As depicted in Figure \textcolor{cyan}{\ref{fig:5}}, the general process of protocol fuzzing comprises four stages: protocol syntax acquisition and modeling, test case generation, test execution and monitoring, and feedback information acquisition and utilization. Each stage in the model encapsulates key principles of protocol fuzzing techniques.

\paragraph{\indent\setlength{\parindent}{1em} \textbf{1) Protocol Syntax Acquisition and Modeling}}
This stage is a critical step before the formal initiation of fuzzing, highlighting the highly structured message syntax and stateful characteristics of the protocol. As indicated by \uppercase\expandafter{\romannumeral 1} in Figure \textcolor{cyan}{\ref{fig:5}}, this stage constructs strict constraints on the protocol input space through information sources such as network traffic, protocol specification documents, or protocol source code. We abstract these constraints as \textbf{protocol message syntax} and \textbf{protocol state syntax}. This stage is intricately connected with other components of modern protocol fuzzers and forms the foundation supporting the entire protocol fuzzing process. Importantly, this stage can run in parallel with the main fuzzing process, dynamically updating as the fuzzing progresses. Various approaches to syntax acquisition and modeling are detailed in section \textcolor{cyan}{\ref{syntax}}.

\paragraph{\indent\setlength{\parindent}{1em} \textbf{2) Test Case Generation}}
This stage focuses on generating specific inputs for the PUT. As depicted by \uppercase\expandafter{\romannumeral 2} in Figure \textcolor{cyan}{\ref{fig:5}}, based on the acquired protocol syntax model and input generation strategy, this stage determines the specific test cases to be provided to the PUT in the fuzzing iteration. Different generation strategies include syntax-based strategies, mutation-based strategies, and the Fuzzer-In-The-Middle (FITM) strategy, which will be discussed in detail in section \textcolor{cyan}{\ref{testcase}}.

\paragraph{\indent\setlength{\parindent}{1em} \textbf{3) Test Execution and Monitor}}
This stage involves injecting and executing the generated test cases and monitoring the protocol program's behavior during execution. As depicted by \uppercase\expandafter{\romannumeral 3} in Figure \textcolor{cyan}{\ref{fig:5}}, due to the vast state space of the protocol, testing deep protocol states requires sending multiple prefix states, incurring significant time overhead. Additionally, the asynchronous nature of network communication brings various time delays, making network sockets significantly slower than traditional file reading. This poses a considerable performance overhead to the network protocol fuzzers and has become a focal point for many research efforts (\textcolor{cyan}{\citep{stateafl}}; \textcolor{cyan}{\citep{11}}; \textcolor{cyan}{\citep{13}}; \textcolor{cyan}{\citep{14}}). Techniques such as virtual machine-level or process-level snapshots, network interface emulation, and I/O synchronization points have elevated the efficiency and throughput of protocol fuzzing to a new scale. These optimizations will be discussed in detail in section \textcolor{cyan}{\ref{execution}}.

\paragraph{\indent\setlength{\parindent}{1em} \textbf{4) Feedback Information Acquisition and Utilization}}
This stage includes the acquisition of feedback information and the utilization of feedback information to guide fuzzing. As depicted by \uppercase\expandafter{\romannumeral 4} in Figure \textcolor{cyan}{\ref{fig:5}}, advanced smart fuzzers are often capable of obtaining feedback information generated during the execution of the target to track the effects of inputs and dynamically guide the entire fuzzing process. The widespread adoption and considerable success of coverage-based greybox fuzzing (CGF) have had a profound impact on fuzzing techniques, demonstrating the effectiveness of coverage information feedback (\textcolor{cyan}{\citep{stateafl}}; \textcolor{cyan}{\citep{11}}; \textcolor{cyan}{\citep{12}}). In section \textcolor{cyan}{\ref{Feedback Information Acquisition and Utilization}}, we summarize the types of feedback information available in the field of protocol fuzzing, including coverage, response codes, state variables, memory, and methods for leveraging this feedback information when fuzzing network protocols.

We argue that the current state-of-the-art protocol fuzzing techniques involve innovation and improvement across the four aforementioned components. This article, based on this foundation, categorizes and compares representative protocol fuzzing techniques, highlighting their advancements in different stages of the unified process model, as shown in Table 2. In many cases, improvements in these four stages are orthogonal. For instance, while proposing more efficient test case generation techniques, further optimization of the execution speed of network protocol fuzzing can be considered.

\begin{table*}
\label{table2}
\centering
\renewcommand{\arraystretch}{1.5}
\resizebox{\linewidth}{!}{

\begin{tabular}{llcccccccc}

\multicolumn{10}{l}{\textbf{Table 2}}     \\

\multicolumn{10}{p{22cm}}{Summary of representative protocol fuzzers and their methodology. This paper distinguishes the key methodological contributions of different fuzzers according to the four execution stages of the general protocol fuzzing process.}  \\

\multicolumn{10}{p{22cm}}{\textbf{general:} This fuzzer can test generic protocols rather than being tailored to a specific one. \textbf{SIP:} Session Initiation Protocol. \textbf{FTP:} File Transfer Protocol. \textbf{WEBAPP:} Web Application. \textbf{SEC:} Security protocols. \textbf{TLS:} Transport Layer Security Protocol. \textbf{ICS:} Industrial Control System Protocol. \textbf{IoTD:} Internet of Thing Devices. \textbf{DTLS:} Datagram Transport Layer Security. \textbf{PAV:} Protocols in Autonomous Vehicles. }  \\

\multicolumn{10}{p{22cm}}{ \textbf{RT Info.}: Runtime Information. \CIRCLE: Blackbox Fuzzer. \LEFTcircle: Greybox Fuzzer. }  \\

\multicolumn{10}{p{22cm}}{\textbf{M:} Message Syntax, representing that the fuzzer needs to manually obtain or be able to self-learn the protocol message format. \textbf{S}: State Syntax, representing that the fuzzer needs to manually obtain or be able to self-learn the protocol state machine and state transition. \textbf{$\Delta M$:} Learn new protocol syntax information during the fuzzing process. }  \\

\multicolumn{10}{p{22cm}}{\textbf{Mut-based}: Mutation-based Protocol Fuzzer. \textbf{Gene-based}: Generation-based Protocol Fuzzer. \textbf{FITM}: Fuzzer-in-the-middle. }  \\

\multicolumn{10}{p{22cm}}{\textbf{SNAP}: Snapshot. \textbf{FuncS}: Network Function Simulation. \textbf{N}: No execution optimization. \textbf{MemFS:} In-memory Filesystem. \textbf{SynP:} I/O Synchronisation Points. }  \\

\multicolumn{10}{p{22cm}}{\textbf{Cov}: Code Coverage. \textbf{S-Cov:} 
Protocol state coverage. \textbf{RC}: Response Code. \textbf{Var:} Branch and Variable. \textbf{Mem}: Memory. \textbf{FC:} Function Code (i.e., A key type of protocol message field in the ICS protocol). \textbf{N:} No feedback information utilization. }  \\

\specialrule{0.1em}{3pt}{3pt} 

\multicolumn{1}{l}{\multirow{2}{*}{Protocol Fuzzer}} &
  \multicolumn{1}{l}{\multirow{2}{*}{Target}} &
  \multicolumn{1}{l}{\multirow{2}{*}{RT Info.}} &
  \multicolumn{2}{c}{1.Protocol Syntax} &
  \multicolumn{3}{c}{2.Testcase Generation} &
  \multicolumn{1}{c}{\multirow{2}{*}{3.Exec-Opt}} &
  \multicolumn{1}{c}{\multirow{2}{*}{4.Feedback}} \\ \cline{4-8} 
\multicolumn{1}{c}{} &
  \multicolumn{1}{c}{} &
  \multicolumn{1}{c}{} &
  \multicolumn{1}{c}{Manual} &
  \multicolumn{1}{c}{Self-learning} &
  \multicolumn{1}{c}{\textbf{Mut}-based} &
  \multicolumn{1}{c}{\textbf{Gene}-based} &
  \multicolumn{1}{c}{FITM} &
  
   & \\ \specialrule{0.1em}{3pt}{3pt} 
PROTOS\citenew{63}     & general & \CIRCLE & M &  &  & \usym{1F5F8} &  & N &  N  \\
Peach\citenew{15}         & general & \CIRCLE & M+S &  &  & \usym{1F5F8} &  & N & N   \\
SNOOZE\citenew{3}        & general & \CIRCLE & M+S &  &  & \usym{1F5F8} &  &  N & N   \\
Sulley\citenew{64}        & general & \CIRCLE & M &  &  & \usym{1F5F8} &  & N & N   \\
Kif\citenew{Kif}           & SIP     & \CIRCLE & M & S &  & \usym{1F5F8} &  & N & N   \\
LZFuzz\citenew{lzfuzz}        & general & \CIRCLE &  & M &  &  & \usym{1F5F8} & N & N  \\
AutoFuzz\citenew{46}      & FTP     & \CIRCLE  &  & M+S &  &  & \usym{1F5F8} & N &  N  \\
AspFuzz\citenew{aspfuzz}       & general & \CIRCLE & M &  &  & \usym{1F5F8} &  & N & N   \\
KFuzz\citenew{kameleonfuzz} & WEBAPP  & \CIRCLE &  & M &  & \usym{1F5F8} &  & N & N   \\
SECFuzz\citenew{70}       & SEC     & \CIRCLE & M &  &  &  &  \usym{1F5F8} & N & N   \\
BooFuzz\citenew{boofuzz}       & general & \CIRCLE & M+S &  &  & \usym{1F5F8} &  & N & N   \\
Pulsar\citenew{45}        & general & \CIRCLE & M+S &  &  & \usym{1F5F8} &  & N & S-Cov  \\
TLSfuzzer\citenew{40}     & TLS     & \CIRCLE & M & S &  & \usym{1F5F8} &  & N & RC  \\
AFLNet\citenew{18}        & general & \CIRCLE+\LEFTcircle &  & S & \usym{1F5F8} &  &  & N & Cov+RC   \\
GANFuzz\citenew{35}       & ICS     & \CIRCLE &  & M &  & \usym{1F5F8} &  & N & N   \\
IoTFuzzer\citenew{iotfuzzer}     & IoTD    & \CIRCLE &  & M & \usym{1F5F8} &  &  & N & N   \\
Polar\citenew{polar}         & ICS     & \LEFTcircle &  & M & \usym{1F5F8} &  &  & N &  Cov+FC  \\
SeqFuzzer\citenew{36}       & ICS     & \CIRCLE &  & S+M &  &  \usym{1F5F8} &  & N & N   \\
Peach*\citenew{66}        & ICS     & \LEFTcircle & M & $\Delta M$ &  & \usym{1F5F8} &  & N & Cov   \\
DTLFfuzzer\citenew{44}    & DTLS    & \CIRCLE & M & S &  & \usym{1F5F8} &  & N & N   \\
Diane\citenew{diane}         & IoTD    & \CIRCLE &  & M & \usym{1F5F8} &  &  & N & N   \\
SGFuzz\citenew{73}        & general & \LEFTcircle &  & S & \usym{1F5F8} &  &  & N & Cov+S-Cov   \\
PAVFuzz\citenew{76}       & PAV     & \LEFTcircle & M & $\Delta M$ &  & \usym{1F5F8} &  & N & Cov   \\
Snipuzz\citenew{37}       & IoTD    & \CIRCLE &  & M & \usym{1F5F8} &  &  & N & N   \\
Nyx-Net\citenew{11}       & general & \LEFTcircle &  & S & \usym{1F5F8} &  &  & SNAP+FuncS & Cov   \\
StateAFL\citenew{stateafl}      & general & \LEFTcircle &  & S & \usym{1F5F8} &  &  & N & Cov+Mem   \\
SNPSFuzzer\citenew{13}    & general & \LEFTcircle &  & S & \usym{1F5F8} &  &  & SNAP & Cov+RC   \\
SNAPFuzz\citenew{14}      & general & \LEFTcircle &  & S & \usym{1F5F8} &  &  & FuncS+MemFS &  N  \\
FitM\citenew{69}          & general & \LEFTcircle &  & S &  &  & \usym{1F5F8} & SNAP+FuncS & Cov+S-Cov   \\
BLEEM\citenew{4}         &    general     & \CIRCLE &  & M+S & \usym{1F5F8} &  &  & N & S-Cov   \\
NSFuzz\citenew{81}        &   general      & \LEFTcircle &  & S & \usym{1F5F8} &  &  & SynP & Cov+RC  \\ 
\specialrule{0.1em}{3pt}{3pt} 
\end{tabular}

}
\end{table*}

\subsection{Categorization of Protocol Fuzzers}
The commonly used classification method for fuzzers is based on the amount of valuable information that can be captured during the execution process (\textcolor{cyan}{\citep{li2018fuzzing}}; \textcolor{cyan}{\citep{yun2022fuzzing}}; \textcolor{cyan}{\citep{1}; \textcolor{cyan}{\citep{58}}}). Following this approach, fuzzers are categorized into three types: blackbox, whitebox, and greybox. The following provides detailed explanations of these three types of fuzzers:

\paragraph{\indent\setlength{\parindent}{1em} \textbf{Blackbox fuzzer}}
Blackbox fuzzers only consider the input and output of the target program, treating it as a completely closed black box and conducting fuzzing without knowledge of the internal implementation details \citenew{godefroid2007random}. It explores vulnerabilities in the program by randomly mutating given initial seeds (e.g., file formats) or generating inputs based on provided input specifications. The effectiveness of a black-box fuzzer during the fuzzing process relies crucially on the quality of the initial seeds or the precision of the input specifications. High-quality seeds and input specifications can explore more program paths in a given time and are more likely to trigger program vulnerabilities. It is worth noting that, unlike conventional software programs, blackbox fuzzing in the network protocol domain can leverage feedback from the target protocol software, such as response codes in reply messages \citenew{18}.
\paragraph{\indent\setlength{\parindent}{1em} \textbf{Whitebox fuzzer}}
Whitebox fuzzers explore the internal structure and branches of a program by accessing its source code or binary files. Based on the code structure, they model the program's control flow, data flow, and other program information (\textcolor{cyan}{\citep{white1}}; \textcolor{cyan}{\citep{white2}}). Utilizing constraint solvers, whitebox fuzzers construct inputs that cover specific branches and analyze whether the program exhibits abnormalities. Compared to blackbox fuzzing, whitebox fuzzing achieves higher code coverage, generates higher-quality test cases, and finds more profound program vulnerabilities. However, in certain specialized scenarios, such as industrial control systems where many protocols are proprietary and their specifications and program source code are inaccessible, the applicability of whitebox fuzzing is significantly limited.

\paragraph{\indent\setlength{\parindent}{1em} \textbf{Greybox fuzzer}}
Greybox fuzzers lie between blackbox and whitebox fuzzers, having access to some portion of the system's source code and internal information. They combine the advantages of both blackbox and whitebox approaches, avoiding the "blindness" of blackbox testing while not requiring access to all information as in whitebox testing (\textcolor{cyan}{\citep{grey1}}; \textcolor{cyan}{\citep{grey2}}). The concept is to provide inputs to the target software, collect valuable feedback during its runtime (e.g., branch coverage), and use this gathered feedback to guide the generation of more effective test cases. Greybox fuzzers can dynamically adjust inputs during the fuzz testing process, enhancing testing efficiency. While greybox fuzzing incurs lower resource overhead compared to whitebox testing, it cannot entirely replace whitebox fuzz testing. In software testing scenarios that require comprehensive testing of internal implementations, greybox fuzzers may have blind spots. Similarly, though more efficient than blackbox fuzzers, greybox fuzzers might be inadequate in certain situations (e.g., industrial control protocols) where access to internal program information is not possible.

Interaction between network protocol software is achieved through protocol messages. Unlike traditional software, in a blackbox scenario, response messages contained within network packets can also serve as feedback to guide fuzz testing \citenew{4}. This distinction makes network protocol fuzzers not entirely conform to the classification characteristics of traditional fuzzers. In practice, the specification for each type of protocol defines the structure of protocol messages and the logic of state transitions. Once the protocol specification is available, a fuzzer can effectively construct request messages based on that specification to trigger different protocol states, enabling more profound fuzzing of network protocol software. However, in certain scenarios, such as private protocols, where the protocol specifications are not publicly available, fuzzers need additional efforts to infer information about the protocol. In light of the characteristics of network protocols mentioned above and based on whether providing the protocol specification is required, we categorize existing network protocol fuzzers:
\paragraph{\indent\setlength{\parindent}{1em} \textbf{Specification-dependent protocol fuzzers}}
We observe that many existing protocol fuzzers require testers to provide a specific protocol specification. Subsequently, these fuzzers generate or mutate new test cases based on this specification. Peach \citenew{15}, for example, needs access to the specification of the protocol under test and uses XML language to write corresponding configuration files. These files include message formats, protocol behaviors, protocol states, and transition relationships between states. Similarly, Peach* and PAVFuzz parse XML-formatted files, converting them into protocol state machine models to generate well-formed messages. Peach* \citenew{66}, an extension of the Peach framework, introduces coverage-guided packet disassembly and generation. It saves packets that trigger new path coverage, breaks them down through the disassembly process into individual fragments, and uses these fragments to construct higher-quality packets for a new round of fuzz testing. PAVFuzz \citenew{76} models the structure of packets based on the protocol specification and introduces a state-sensitive mutation strategy to enhance the fuzz testing efficiency of Peach. Since users provide information about the tested protocol's specifications, these fuzzers are typically more efficient during testing and more likely to uncover deeper states and vulnerabilities in the tested protocol. However, in recent years, with the application of private protocols, especially in the industrial control protocol domain, where protocol specifications are not publicly available, specification-dependent protocol fuzzers cannot achieve ideal testing results in such scenarios.

\paragraph{\indent\setlength{\parindent}{1em} \textbf{Specification-free protocol fuzzers}}
Due to the specificity of network protocols, especially private protocols used in industrial control systems, many protocol specifications are unavailable. For such protocols, testers can only infer the program's state and the required format of input message sequences at runtime by capturing network traffic or utilizing feedback from the program's execution (such as response codes). Typically, traffic transmitted between protocol software is done through unique combinations of source IP, destination IP, and ports. PULSAR \citenew{45} integrates the concepts of fuzzing with automatic protocol reverse engineering and simulation techniques. It captures, identifies, and extracts all traffic transmitted between protocol software, ultimately inferring message formats and the state machine model of the protocol through methods like clustering. PULSAR generates test cases based on this model, simulating the flow of traffic in subsequent fuzz testing processes to discover deep vulnerabilities in protocol software. Inspired by man-in-the-middle attacks (see Section \textcolor{cyan}{\ref{FITM}}), AutoFuzz \citenew{46} learns protocol implementations by constructing a finite state machine (FSM). The FSM automatically captures the communication flow between the client and server. Additionally, AutoFuzz employs bioinformatics algorithms to determine fields within individual protocol messages, categorizing them as fixed or variable. It intelligently modifies communication between the client and server for fuzz testing under the guidance of the FSM. Even without access to the protocol specification, specification-free protocol fuzzers can model the syntax of the testing protocol proficiently using techniques such as protocol reverse engineering and man-in-the-middle attacks. In recent years, this approach has gained significant traction, especially in private protocols.

\section{Protocol Syntax Acquisition and Modelling}
\label{syntax}

Protocol syntax acquisition and modeling are crucial stages that many modern network protocol fuzzers need to perform before they can formally enter fuzzing. Compared to traditional fuzzing objects such as file handlers (\textcolor{cyan}{\citep{19}}; \textcolor{cyan}{\citep{20}}), protocol software has more complex constraints on the effective input space. We abstract the numerous constraints on protocol inputs at a high level into two syntaxes: protocol message syntax and protocol state syntax. The purpose of this stage is precisely to complete the inference and acquisition of the above two syntaxes to provide key information for subsequent syntax-based fuzzing. Effective protocol syntax acquisition can greatly reduce the gap between the input space and the effective input space, taking the efficiency of fuzzing to a new dimension \citenew{1}. It is worth noting that some fuzzers can reason during the fuzzing process in conjunction with program feedback and dynamically update the protocol syntax model (\textcolor{cyan}{\citep{11}}; \textcolor{cyan}{\citep{16}}; \textcolor{cyan}{\citep{17}}; \textcolor{cyan}{\citep{18}}).

\subsection{Terminologies}

\subsubsection{Network Protocol Syntax}
Protocol syntax is part of network protocols and is designed to describe the structure, format, encoding method, and some other information about the data to transmit it securely and correctly to the receiver. Protocol syntax can be divided into two types: message syntax and state syntax. Message syntax is used to transmit data and metadata in communication, such as the content and size of the file being transferred. State syntax is used to describe the state of the system during communication. In stateful network protocols, the two together determine the effective input space and the desired output space at any point in time throughout the runtime cycle of the protocol software. 

\begin{figure}[h]
    \centering
    \includegraphics[width=0.48 \textwidth]{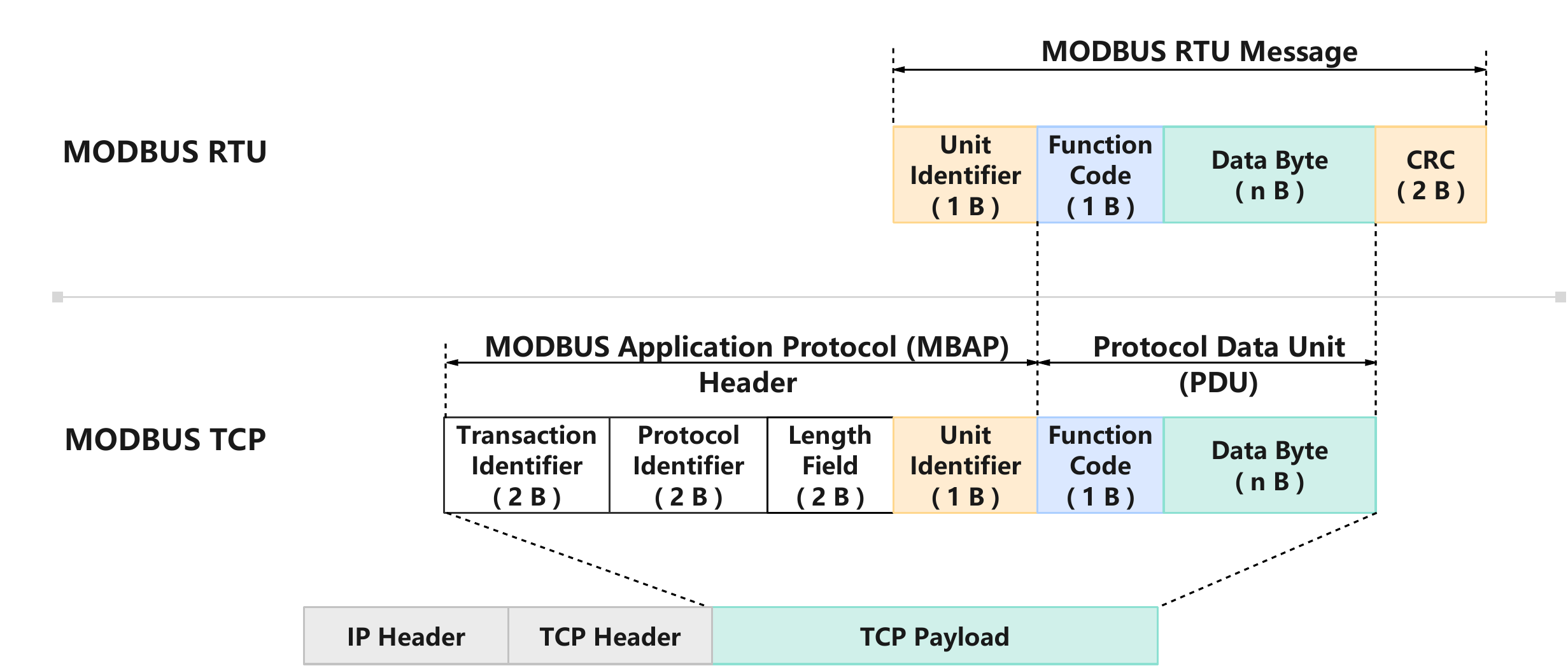}
    \caption{Modbus Protocol Specification.}
    \label{fig:6}
\end{figure}

\paragraph{\indent\setlength{\parindent}{1em} \textbf{Message Syntax}}
Message syntax is a regularised encoding method for describing the structure and syntax rules of valid messages in a protocol without being influenced by context, environment, or previous messages. In other words, message syntax provides a standardized format to ensure correct and reliable communication. Message syntax specifies the division of the different fields in a message, the order in which the fields are organized, and the dependencies between fields (e.g., the contents of some fields are the length and checksum of other fields). Within each field are attributes such as data type, length, and value range. As shown in Figure \textcolor{cyan}{\ref{fig:6}}, the Modbus protocol specification divides the message into fields such as transaction identifier, protocol identifier, length, and data fields, each with its own specified concrete value or valid value interval, length, and other attributes. 

\paragraph{\indent\setlength{\parindent}{1em} \textbf{State Syntax}}
State syntax specifies the set of states of a network protocol and the logic of transitions between states, usually described by protocol state machines. As shown in Figure \textcolor{cyan}{\ref{fig:7}}, state syntax is a detailed description of the set S of states and the set t of state transition logic. The client must send messages in a specific sequential order to interact correctly with the server. 

\subsubsection{Protocol Specification}
Protocol specifications are usually provided in the form of an official document that includes a complete description of the message syntax and state syntax as well as some other relevant information such as protocol architecture, error handling, security considerations, and extensibility. A common category of protocol specification is RFC \citenew{21}, e.g., the FTP protocol is defined by RFC 959, 3659, 2228, 2428, 5797. 

\subsubsection{Network Protocol Reverse Engineering}
Network protocol reverses engineering \citenew{26} refers to the automated process of extracting protocol format, syntax, and semantics by monitoring and analyzing the inputs and outputs of the protocol software, the system behavior, and the instruction execution flow without relying on the protocol specification. It is characterized by its a priori-free nature, i.e., it does not require prior knowledge of the protocol specification.

\subsection{Message Syntax Acquisition}
\label{32}
Network protocol fuzzing is highly dependent on message syntax. For most syntax-based network protocol fuzzers, message syntax is essential (\textcolor{cyan}{\citep{15}}; \textcolor{cyan}{\citep{22}}). This section systematically summarises the methods for acquiring protocol message syntax, including both manual acquisition and automatic learning. It is worth noting that some of the work in this section (e.g., automatic learning) is not a direct improvement on fuzzing algorithms or techniques, but is more oriented towards protocol reverse aspects (\textcolor{cyan}{\citep{23}}; \textcolor{cyan}{\citep{24}}). For instance, techniques based on field segmentation, keyword recognition, and delimiter recognition. These techniques extract the protocol message format and then infer the semantics in context. 

\begin{figure*}[h]
    \centering
    \includegraphics[width=0.98 \textwidth]{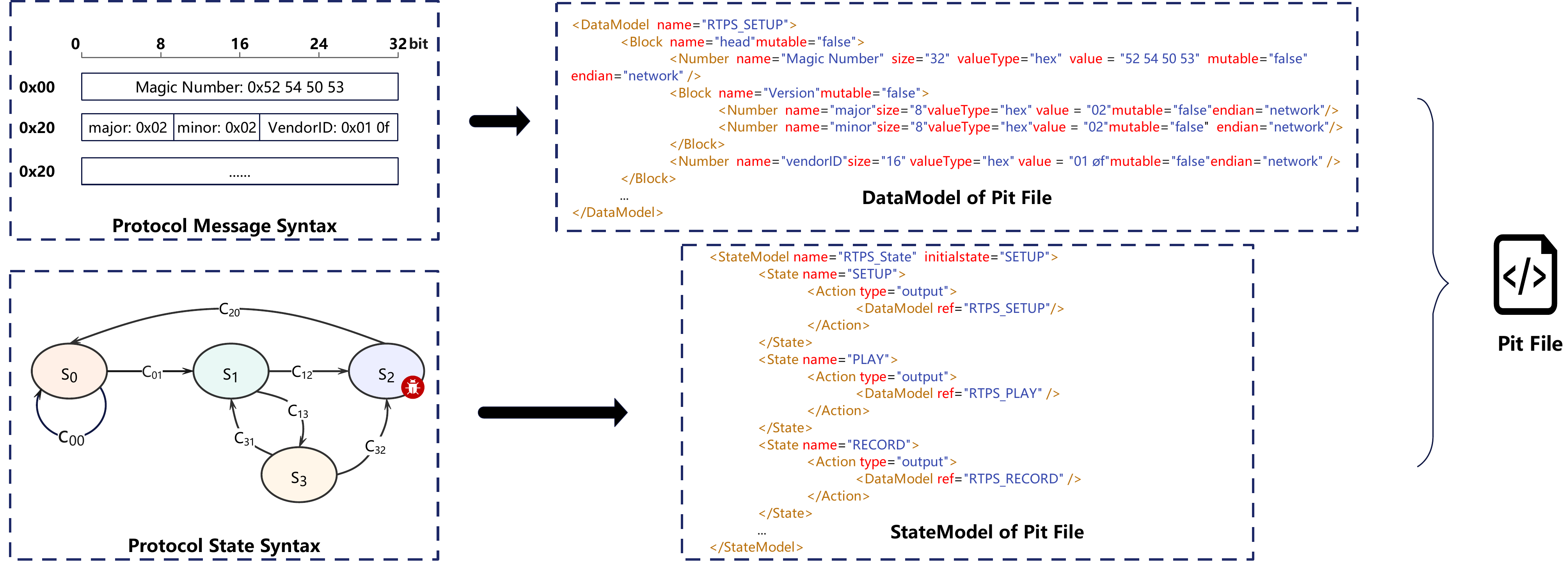}
    \caption{Peach Pit File.}
    \label{fig:7}
\end{figure*}

\subsubsection{Manual Acquisition}
Manual extraction of protocol syntax from official protocol documentation, source code, and captured network traffic is the most straightforward way. In most of the current mainstream protocol fuzzers, manual acquisition of protocol message syntax is required, e.g., Peach \citenew{15}. The premise of using Peach is that the user can write a Pit file using the XML language while being familiar with the target protocol syntax. As shown in Figure \textcolor{cyan}{\ref{fig:7}}, the DataModel of the file is a description of the protocol syntax. 

However, manual protocol syntax extraction is a tedious, time-consuming, and error-prone process. More seriously, 1) manual extraction methods are not scalable, and when confronted with a completely new protocol, all the work has to be completely restarted; 2) Currently, a large amount of protocol security analysis work involves objects that do not have publicly available specifications or description files, such as industrial control systems (ICS). It uses private protocols due to their specificity, which makes manual extraction of message syntax almost impossible to achieve.

\subsubsection{Automatic Learning}
Automatic learning of protocol specifications involves the use of techniques such as traffic analysis and deep learning. The fuzzers use machine learning techniques to analyze the response of the protocol to various input behaviors. Based on the information obtained, the fuzzer can automatically generate a formal specification of the protocol for further testing and validation. Some fuzzers use clustering or classification methods \citenew{37} to gather behaviors generated by different inputs according to the procedure and identify patterns in the protocol behaviors. The fuzzer can use these patterns to generate a formal syntax specification for the target protocol, improving the accuracy and efficiency of fuzzing. Specifically, the following approaches exist:
\paragraph{\indent\setlength{\parindent}{1em} \textbf{Network Traffic Analysis}}
Network traffic analysis refers to building a corpus by capturing network traffic during the normal operation of a protocol and extracting from it different fields, delimiters, and other message syntax of the protocol by combining various heuristic algorithms. For instance, Discoverer \citenew{26} utilizes many idioms that are common in application layer protocols and extracts syntax information from them by clustering network traffic. Beddoe \citenew{67} utilizes bioinformatics algorithms to analyze consistent sequences samples by aligning similar message; Roleplayer \citenew{25} uses a byte-stream alignment algorithm to identify fields that can be mutated to abstract protocol specifications; IPspex \citenew{27} uses field semantics to locate the target message of an ICS protocol, and then uses a data-stream backtracking and sequence-alignment algorithms to infer message syntax. Network traffic analyses are usually more effective in practice, but are often limited by the accuracy of the reasoning \citenew{22}.
\paragraph{\indent\setlength{\parindent}{1em} \textbf{Procedural Behaviour Analysis}}
Program behavior analysis refers to the whitebox approach to analyze the specific behavior of a program when processing message data, thus obtaining a large amount of information about the protocol message format (\textcolor{cyan}{\citep{31}}; \textcolor{cyan}{\citep{32}}; \textcolor{cyan}{\citep{33}}; \textcolor{cyan}{\citep{34}}). Polyglot \citenew{30} is a pioneer in the use of dynamic analysis techniques (e.g., taint analysis and symbolic execution), which successfully extracted the message syntax of protocols from records of program behavior by monitoring the actions of programs as they processed input. Polar \citenew{polar} combines static analysis with byte-level dynamic taint analysis to extract protocol function code variables and their associated message syntax units. NSFuzz \citenew{81} also uses static analysis to identify and filter state variables in protocol source code and event loops during protocol execution. However, this approach cannot be used in blackbox scenarios where protocol binaries and source code are not accessible.
\paragraph{\indent\setlength{\parindent}{1em} \textbf{AI-assisted learning methods}}
In recent years, machine learning methods have contributed to several aspects of fuzzing. Among them, Natural Language Processing (NLP) techniques are widely used for protocol grammar analysis. Several studies have implemented automatic protocol message syntax learning systems based on this (\textcolor{cyan}{\citep{34}}; \textcolor{cyan}{\citep{35}}; \textcolor{cyan}{\citep{36}}).GANFuzz \citenew{35} proposes a specification-free testcase generation method for ICS protocol. It learns the protocol syntax by training a generative model in a generation-adversarial network to estimate the latent distribution function of protocol messages. However, these methods are expensive to train and the trained protocol specifications are greatly limited by the set of test cases. More importantly, when dealing with complex scenarios, NLP models often struggle to predict and explain program behavior.
\paragraph{\indent\setlength{\parindent}{1em} \textbf{Message Fragment Inference Method}}
Snipuzz \citenew{37} proposed a novel approach: mutate the message to be sent byte by byte and collect the responses generated by the mutated message, combining the bytes that triggered the same response into a fragment based on the consistency of the obtained responses, with each fragment corresponding to a specific code execution path in the protocol. Although it may differ from the actual syntax rules, Snipuzz builds a hidden, functionally oriented special message syntax structure that fuzzers can use to efficiently constrain the input space. However, for some complex applications and protocols, this approach may not be able to determine the correct message structure and content. In addition, a large amount of redundant test cases are generated during the message field inference process, resulting in a waste of resources.

\subsection{State Syntax Acquisition}
\label{33}
The input space of a protocol is constrained by message syntax and state syntax. State syntax focuses on describing the dynamic properties of the protocol software, which need to be obtained by analyzing the protocol behavior. The results of protocol behavior analysis are generally obtained by constructing a protocol state machine, which mainly includes the steps of protocol session segmentation, different types of message sequence construction, and state machine concise. State syntax-based guided fuzzers send a specific sequence of messages in the protocol state space to reach the target state of interest. Although part of the mutation-based fuzzers do not consider the message syntax format, they need to consider the protocol state syntax when fuzzing protocols. This section systematically summarises the three basic approaches currently used to infer protocol state syntax, i.e., manual learning, active inference, and passive inference.

\subsubsection{Manual Learning}
Manual learning of state syntax involves manually analyzing the state transitions of a network protocol and then writing syntax rules to describe these transitions. These syntax rules specify the inputs and outputs in network protocol state transitions, as well as changes in protocol state. This is a customized approach for a specific network protocol and requires an in-depth understanding of how the protocol works and the details of state transitions. The general idea is essentially the same as manually acquiring message syntax. The main difference between the two is that the data structure describing the state syntax is different from that describing the message syntax, and state syntax is usually described by constructing a finite state machine (FSM), as shown in the DataModel of Pit File in Figure \textcolor{cyan}{\ref{fig:7}}. In the Pit file, the StateModel module is utilized to describe the protocol state and state transition logic from the fuzzer's perspective, as shown in the StateModel of Pit File in Figure \textcolor{cyan}{\ref{fig:7}}. By learning the state syntax manually, testers can build more efficient fuzzing test cases to help discover vulnerabilities and security issues in network protocol implementations. 

\subsubsection{Active Inference}
Active inference is a stage independent of the main process of fuzzing testing. In this phase, the learning machine actively generates a series of test messages and sends them to the PUT to obtain the corresponding outputs, while using model learning algorithms (e.g., Angluin's L* algorithm \citenew{38}, LearnLib \citenew{39}) to infer and construct the complete state machine of the target protocol. Through this series of work (\textcolor{cyan}{\citep{40}}; \textcolor{cyan}{\citep{41}}; \textcolor{cyan}{\citep{42}}; \textcolor{cyan}{\citep{43}}; \textcolor{cyan}{\citep{48}}), fuzzers can obtain a protocol state syntax and perform stateful fuzzing on target protocol software based on that syntax. For instance, TLS-Attacker pioneered the use of protocol state fuzzing (or simple stateful fuzzing). Rather than looking for defects that traditional fuzzing is good at finding, such as buffer overflows, string error handling, etc., the purpose of stateful fuzzing is to search for state machine bugs. Stateful protocol fuzzers can utilize complex protocol state machines constructed by active learning to test for flaws in state transition logic that can be exploited to construct attacks such as authentication bypass \citenew{44}. The active inference approach applies to protocols where the state space is not very complex.

\subsubsection{Passive Inference}
Different from active inference, passive inference does not require active messages to be sent to the test object to accomplish state learning before the fuzzing formally starts. Passive inference learns state transition logic through samples (\textcolor{cyan}{\citep{45}}; \textcolor{cyan}{\citep{46}}; \textcolor{cyan}{\citep{47}}). There are two main sources of samples: network traffic generated by the protocol software under normal behavior and test cases used during fuzzing performed in parallel. AFLNet \citenew{18} dynamically maintains the finite state machine (FSM) of the protocol by determining whether this test explores a new state of the protocol based on the response code feedback during the fuzzing. Prospex \citenew{22} acquires network traces of protocol runs and abstracts message formats to construct message clustering (message clustering). Based on this, it constructs a state tree using a heuristic algorithm [49] to merge similar states and generate a minimum state machine. The disadvantage of passive inference protocol state syntax is that it is difficult to capture the full context of the protocol state, which makes it more challenging to generate inputs approximating the real situation in subsequent stages.
\section{Test Case Generation}
\label{testcase}
This section will describe in detail methods for generating test cases. In the unified process model shown in Figure \textcolor{cyan}{\ref{fig:5}}, the inputs to the module are the protocol syntax model and the generation strategy, and the outputs are the actual test cases executed by the protocol software (e.g., server, client). In network protocol fuzzing, the test cases are usually sequences of packets in chronological order, which are sent by the fuzzer to the network protocol software under test in a given order and complete execution. The quality of the test cases greatly determines whether or not a vulnerability or security flaw in the program under test will be triggered, so the approach and decisions made at this stage will directly impact the effectiveness of fuzzing.

The approach to generating effective test cases involves two key issues:

Q1:How do generate test cases that can pass protocol message syntax checking?

Q2:How do make the generated message sequences effective in exploring the complex state space of a protocol?

Depending on the test case generation method, fuzzers are usually classified into two categories: generation-based and mutation-based. Generation-based fuzzers use a provided syntax model to generate test cases. Mutation-based fuzzers mutate the seed file provided by the user using various mutation operators, which in turn generates large amounts of test cases. It is worth noting that, inspired by the idea of MITM attacks, in recent years there has been a category of fuzzers that intercept the communication traffic between the client and the server and modify it as a way to construct test cases. This category of fuzzers has shown unique advantages in the field of protocol fuzzing. We refer to such fuzzers as FITM (fuzzer-in-middle). The network topology of the FITM at runtime is different from the traditional case shown in Figure \textcolor{cyan}{\ref{fig:2}}, as it acts as an intermediary to intercept the communication traffic between the client and the server, rather than acting as a server or a client as in the traditional fuzzer idea, as shown in Figure \textcolor{cyan}{\ref{fig:9}}. The following is a detailed description of these three categories of fuzzers.

\subsection{Generation-based Protocol Fuzzers}
Generation-based protocol fuzzers utilize two categories of syntax models (i.e., message syntax and state syntax described in Section 3) to resolve the above two key issues separately. Corresponding to the two categories of syntax, the generation-based protocol fuzzers need to accomplish two dimensions of generation: the generation of individual messages and the combination of message sequences.

The earlier protocol fuzzers such as Peach \citenew{15}, PROTOS \citenew{63}, Sulley \citenew{64}, SPIKE \citenew{65}, and SNOOZE \citenew{3} are typical generation-based fuzzers. These fuzzers provide a standard interface in the form of a configuration file for the user to use, and the user needs to write a test configuration file for the target protocol based on the tool specification and protocol syntax. Configuration files usually contain some form of a data structure describing the protocol syntax, so the process of writing a configuration file is actually a modeling of the protocol syntax from the user's point of view.

Peach is a representative of generation-based fuzzers, which requires the user to write a configuration file, i.e., a Pit file, using the XML language, as shown in Figure \textcolor{cyan}{\ref{fig:7}}. The Pit file consists of five parts: a generic configuration, a data model, a state model, an agent and a monitor, and a fuzzing configuration, where the data model and the state model are syntax modeling of the target protocol. Data models provide sub-elements at different levels of granularity, such as Number, Blob, or String, which are used to define the message syntax structure of a single message. Generally, a complete message is described by a data model. The state model describes the basic state machine logic required to test a protocol via the state and action sub-elements. Peach generates individual messages by selecting several variable data fields from the corresponding data model and uses a mutation operator to generate test cases by randomly mutating these fields; message sequences are generated by executing the local state machine logic defined in the state model. SNOOZE provides the user with a set of fuzzing primitives, and a fuzzing scenario summary document written using the fuzzing primitives defines the complete process of fuzzing. For example, the SnoozeMessage primitive is used to declare a message object, the setField primitive is used to specify the values of data fields in a message, and so on.

This initial generation-based fuzzer can efficiently generate inputs that conform to the protocol specification, guided by the syntax model. However, its dependence on the quality of the syntax model greatly limits the effectiveness of the input. Incomplete or even incorrect syntax models can lead fuzzers to an inefficient dilemma, while manually extracting complex protocol syntax specifications is usually a time-consuming and error-prone process. More importantly, such fuzzers lack the ability of protocol state space heuristic auto-exploration and can only execute exactly according to a pre-set finite state machine, e.g., Peach repeatedly iterates through the execution of all actions defined in the state model.

The current advanced protocol fuzzer improves on several aspects of the test case generation scheme. Different from the probabilistic selection decision of target mutation fields such as Peach, PAVFuzz \citenew{76} argues that data elements in a data packet have different levels of importance. It calculates dynamic mutation weights for data elements and stores them in a relational table, performing state-sensitive mutations on data elements that are more likely to trigger system vulnerabilities. Peach* \citenew{66} utilizes coverage feedback to preserve valuable packets, decompose them into fragments, and construct higher-quality test cases in the next round of fuzzing. In addition, research on protocol syntax self-learning (as we describe in Section 3) can also contribute directly to the efficiency improvement of generation-based fuzzers. Prospex \citenew{22} extends the work on acquiring message formats based on the analysis of program behavior and introduces techniques for identifying and clustering different types of messages. It is capable of automatically inferring state machines and supports the automatic transformation of the two types of syntax obtained from learning into Pit files, which contributes positively to the development and wide application of Peach.

\begin{figure}[h]
    \centering
    \includegraphics[width=0.48\textwidth]{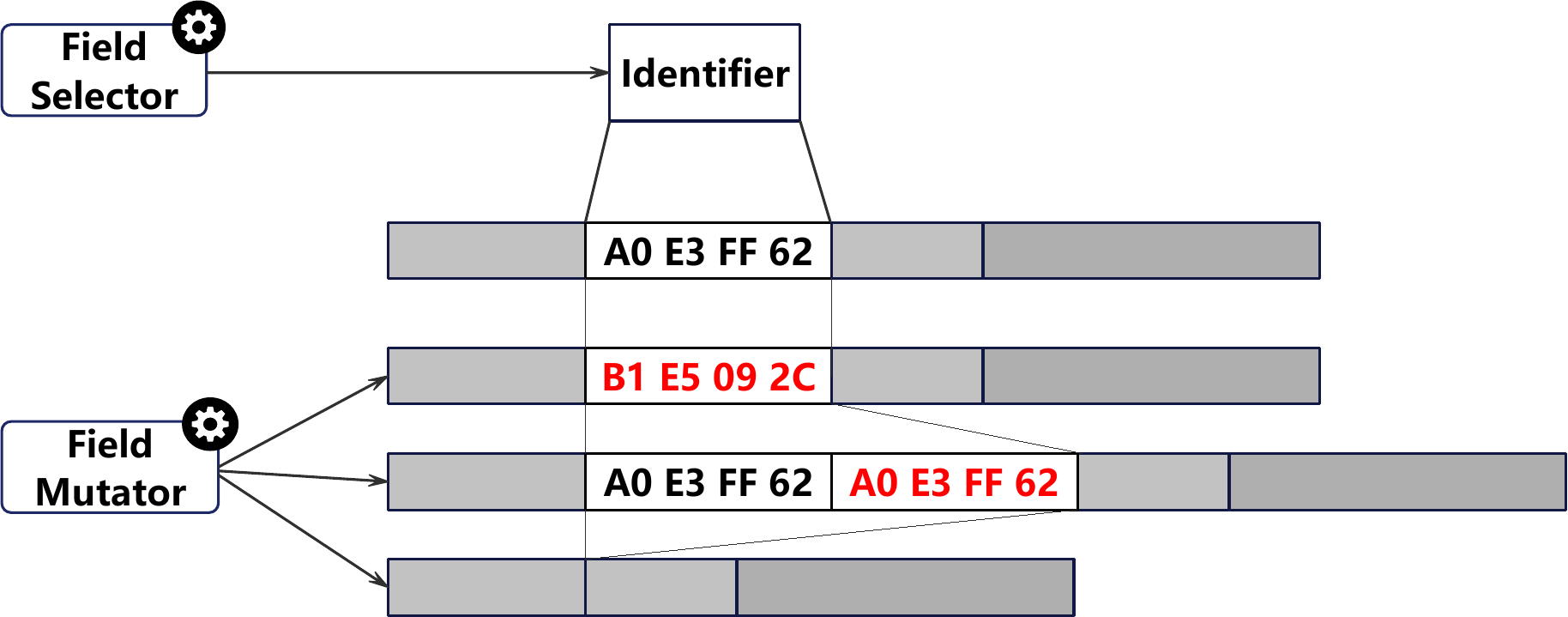}
    \caption{Mutation-Based Fuzzers: Principles of Mutator Operation.}
    \label{fig:8}
\end{figure}

\subsection{Mutation-based Protocol Fuzzers}
The mutation-based protocol fuzzer does not need to provide a syntax specification. It uses various mutation strategies (e.g., bitflip, arithmetic, havoc, and splice) on pre-provided seeds to generate test cases, As shown in Figure \textcolor{cyan}{\ref{fig:8}}. Due to its simplicity and ease of implementation, this category of fuzzers is extensively used. Based on the traditional mutation-based fuzzer, AFL pioneered the introduction of coverage feedback, making coverage-based greybox fuzzer (CGF) one of the hottest research directions in the fuzzing field in recent years. However, the application of mutation-based fuzzers in the field of protocol fuzzing is greatly hampered by the specificity of network protocols compared to traditional software. Without syntax guidance, bit granularity mutation severely disrupts the message structure, making it difficult for test cases to pass the initial syntax check of the protocol. The results of an existing study \citenew{75} showed that up to 90\% of the test cases generated using the AFL mutation strategy were rejected by the protocol software due to syntax errors that failed the initial check.

To break this inherent limitation of traditional fuzzers, advanced mutation-based protocol fuzzers attempt to use protocol syntax to guide the mutation of test cases (\textcolor{cyan}{\citep{11}}; \textcolor{cyan}{\citep{13}}; \textcolor{cyan}{\citep{14}}; \textcolor{cyan}{\citep{18}}; \textcolor{cyan}{\citep{56}}; \textcolor{cyan}{\citep{59}}), i.e., the learned syntax specification is utilized to protect the structure of the inputs to be tested from being corrupted during the mutation process. This attempt also blurs the boundary between mutation and generation and is gradually becoming a tendency. AFLNet sequentially combines all messages sent by a client during a single interaction into a single seed, while using specific ASCII characters to divide the different messages. At the same time, it maintains and updates a finite state machine by extracting response codes. During the fuzzing process, AFLNet can not only judge the quality of test cases through state feedback but also continuously add newly discovered states to the existing state machine, thus completing the self-learning of protocol state syntax. StateAFL \citenew{stateafl} similarly learns the state syntax of the protocol software continuously during the fuzzing process. It uses snapshot techniques to track memory allocation and network IO operations of the PUT and maps them to unique state identifiers. Since many protocols do not embed specific state description fields (e.g., state codes, etc.) in response messages, this state extraction method is more widely used than AFLNet. It is worthy of mentioning that IJON \citenew{50} argues that current fuzzers do not correctly explore the state of a program that exceeds code coverage, e.g., where program execution leads to the same code coverage, but with different state traces. The human analyzer is capable of greatly improving the efficiency and performance of the fuzzer by annotating those parts of the state space that should be explored more thoroughly (usually with only one or two additional lines of code). Therefore, based on AFL, IJON introduces a manual labeling mechanism to guide CGFs to explore the hard-to-cover parts of the state space more efficiently.

In addition to the need to match the protocol syntax to ensure the validity of the generated test cases, mutation-based fuzzing has several inherent disadvantages. For example, the quality of the seed file severely affects the effectiveness of mutation-based fuzzers. In the absence of high-quality, type-rich seed files, the fuzzer is much less likely to find deeper vulnerabilities, and it will take longer to discover them, ultimately leading to relatively low overall efficiency.

\begin{figure}[h]
    \centering
    \includegraphics[width=0.48\textwidth]{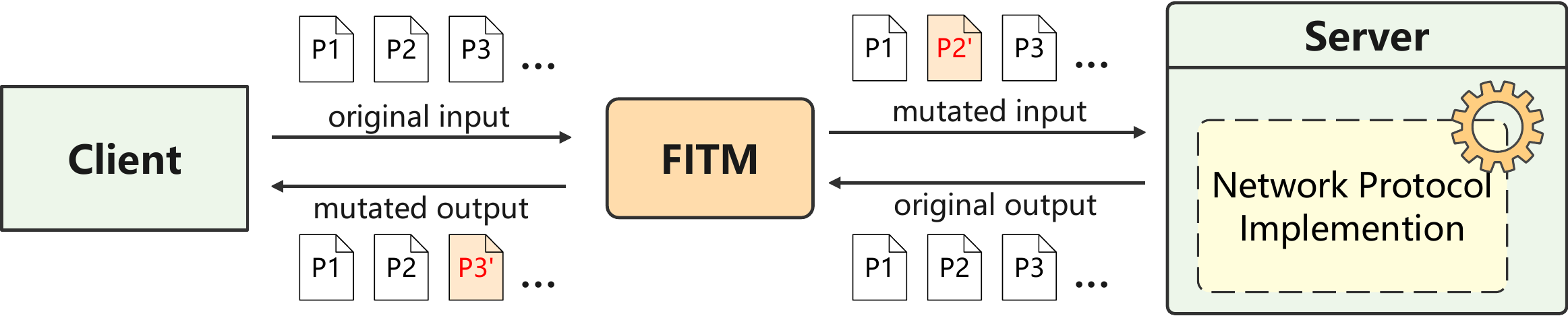}
    \caption{FITM (Fuzzer-in-the-Middle) Fuzzing Process.}
    \label{fig:9}
\end{figure}

\subsection{FITM (fuzzer-in-middle)}
\label{FITM}
Man-in-the-middle (MITM) attack is a method of network attack. An attacker intercepts an existing dialogue or data transmission between two parties by eavesdropping or masquerading as a legitimate participant and unknowingly accessing, tampering with, or manipulating information exchanged between the parties. The protocol fuzzer (man-in-the-middle fuzzer) designed based on this idea captures the network traffic between the server and the client and selectively tampers and replays it as shown in Figure \textcolor{cyan}{\ref{fig:9}}. Such fuzzers do not need to do the complex, syntactically-compliant message construction themselves, and they also alleviate the dependence on state syntax. Notably, FITM can initiate the fuzzing process at any stage of the communication between the client and the server. For example, if a tester wishes to explore vulnerabilities in the protocol implementation at a later stage of the communication, FITM can do this without disrupting packets from earlier communications. Moreover, in the network topology of FITM, the server and the client are two peers, which makes it possible to perform fuzzing for both the client and the server at the same time \citenew{70}. 

In recent years, more and more protocol fuzzers have turned to using man-in-the-middle attacks to generate test cases, e.g., ProxyFuzz \citenew{proxyfuzz}, AutoFuzz \citenew{46}, SECFUZZ \citenew{70}, FITM \citenew{69}, and so on. However, this approach requires specialized tools and techniques to intercept and modify the communication between protocol software, which can increase the complexity of the fuzzing process. Also, in special fuzzing situations, FITM requires as detailed a specification of protocol syntax as other generation-based fuzzing. For example, when performing fuzzing on security protocols where encryption mechanisms are present, the FITM needs to know the encryption and decryption algorithms to perform effective mutations on network traffic.
\section{Test Execution and Monitor}
\label{execution}
Test execution denotes the process in which the PUT receives and executes test cases generated by the fuzzer. The monitor senses the trigger of a bug by monitoring the running state of the PUT. Like common fuzzers, most protocol fuzzers catch abnormal crashes of the program under test or monitor them in combination with memory error-checking tools such as Sanitizer provided by the compilation toolchain.

In fuzzers for common software, test execution is not a worthwhile concern as the execution process is usually very simple. For example, AFL \citenew{12} redirects the standard input of the program to be tested directly to the test case file or adds the file path of the test case in the command line arguments. In the protocol fuzzing scenario, the test execution process imposes a huge performance load on the protocol fuzzer, which seriously impacts the fuzzing efficiency.

First, the protocol software transmits data through the network interface. Specifically, the network interface constructs and sends data according to the protocol, and also receives and interprets data received from the network to transmit it to the higher-level protocol stack. This also means that traditional methods of transmitting input via redirects or command-line arguments are not viable. In the most ideal case, the server process under test is located on the same physical machine as the client process, in which case it is still necessary to use the local loopback IP address as well as the port number, and then complete the input of the test cases over a standard network interface (e.g., TCP or UDP). Since network interfaces are much slower compared to file reads, this dependency results in additional time overhead. Second, complex, multi-threaded server processes generally have a higher startup time cost. Further, since there is no way to confirm that the server program after a new startup or receipt of a message enters a state where it can continue to receive messages, the fuzzer often needs to empirically set a regular waiting time, which introduces another portion of the time overhead. Finally, part of the protocols require extra effort to fully restore to the pre-test state, e.g., the FTP protocol may have changes to the target's filesystem during testing.

The above problems have attracted extensive attention from researchers. In recent years, some advanced works have keenly observed this and proposed some efficient techniques to increase the speed and throughput of protocol fuzzing to a new level of order of magnitude.

\subsection{Snapshots}
Snapshots are static copy files that store the state of the operating system or process in physical memory and various devices at a particular moment. Commonly, there are system snapshots, virtual machine snapshots, file system snapshots, etc. By taking and recovering snapshots, a specific target state can be completely preserved and recovered. Snapshots have a wide range of practical applications in various scenarios. Xu et al. \citenew{52} pioneered the use of snapshotting techniques in fuzzing, providing optimized fuzz primitives replacements for system calls such as fork, speeding up LibFuzzer by up to 736 times. After that, snapshot techniques have been extensively focused and applied (\textcolor{cyan}{\citep{11}}; \textcolor{cyan}{\citep{13}}; \textcolor{cyan}{\citep{14}}; \textcolor{cyan}{\citep{52}}; \textcolor{cyan}{\citep{53}}; \textcolor{cyan}{\citep{54}}; \textcolor{cyan}{\citep{55}}; \textcolor{cyan}{\citep{57}}; \textcolor{cyan}{\citep{69}}), and the application in the field of protocol fuzzing is also one of the successful practices.

In addition to improving the state restoration speed of the PUT in fuzzing, the snapshot technique enables the fuzzer to span the state space of the protocol at a minimal cost. Nyx-net \citenew{11} introduces kernel-state, VM-level incremental snapshots to ensure that all states of the PUT are reset before each test case is executed. SNPSFuzzer \citenew{13} uses the user-state, process-level snapshot tool CRIU to save the context of each state to be fuzzed in the PUT and restores that snapshot directly when fuzzing a specific state is required. This saves a large amount of time wasted during test execution by sending frequent prefix messages to span the state space. FITM \citenew{69} also uses CRIU to independently fuzz each state transition by creating a snapshot of each protocol's new state discovered by the fuzzer.

\subsection{Network Function Replacement or Emulation}
\label{section52}
Some research has focused on improving the underlying causes of inefficiency in protocol fuzzing by replacing network function API, file system API, etc. with more efficient interface functions or customized simulation methods. Nyx-net implements a virtual machine-based emulation of relevant network functions. It uses the hook to intercept network functions, APIs related to file descriptor operations (e.g., \textit{accept(), recv(), dup(), close()}), and a customized emulation API to implement a high-performance and high-throughput conversion interface. SnapFuzz \citenew{14} is enabled to intercept system calls issued by the program under test, replacing slow standard Internet sockets with fast UNIX Domain sockets. In addition, it replaces the operating system's persistent filesystem API with an in-memory filesystem API to mitigate the time overhead caused by filesystem resets.FITM uses hooks instead of slow sockets and shared maps instead of network system calls.

\subsection{Protocol I/O Synchronisation Points}
NSFuzz \citenew{81} proposes a solution to set I/O synchronization points based on protocol event loops to accelerate test execution, based on insights into how code is implemented in the service processing phase of network protocol software. Specifically, the network protocol enters its main service processing stage after completing its initialization work and waits for message input from the client. Service processing is usually implemented as an iterative event loop, where each client input triggers a round of related message processing, and after completing the current processing, it returns to the event loop entry to wait for the next input from the client. Therefore, NSFuzz takes the entry of the event loop as a sign that the server can accept the next input, and sets the IO synchronization point at the loop entry to send the fuzzer a sign that it can continue. This design effectively mitigates the overhead associated with setting a fixed sending time window on the fuzzer side and optimizes the performance of test execution.
\section{Feedback Information Acquisition and Utilization}
\label{Feedback Information Acquisition and Utilization}
Feedback refers to a category of available attributes contained in the output produced by a software system. The fuzzer extracts valid information from these attributes and makes dynamic adjustments to the subsequent fuzzing process, e.g., by making heuristically fuzzing decisions. Passive learning, as mentioned above, involves collecting state traces to infer information about the input format. Different from passive learning, feedback is a form of active learning that interacts with the software program to be tested during fuzzing. The fuzzer tracks and evaluates the effectiveness of existing test cases, using the results of historical test cases to make adjustments before attempting the next input. Without the introduction of feedback and heuristic strategies, fuzzing simply remains an inefficient blind random test.

Since the popularity of coverage-based greybox fuzzing (CGF) represented by AFL \citenew{12}, most of the state-of-the-art fuzzers proposed in academia and industry have been intelligently guided to proceed with subsequent fuzzing tests based on the feedback of the results of previous tests. In this section, we systematically summarise the commonly used categories of feedback in existing protocol fuzzers (including response codes, coverage, branch, variables, and memory) and introduce the corresponding representative fuzzers.

\subsection{Response Code}
\begin{figure}[h]
    \centering
    \includegraphics[width=0.48\textwidth]{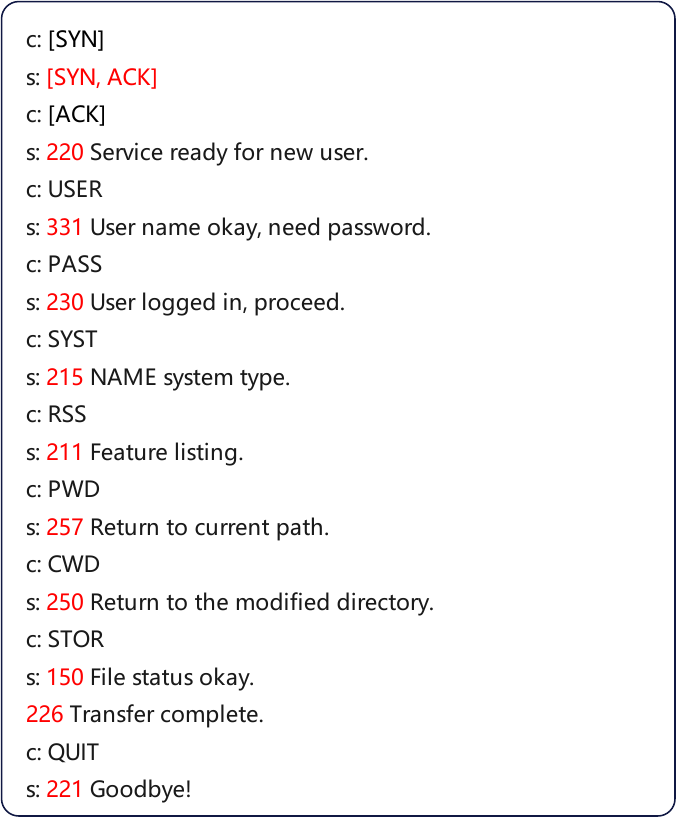}
    \caption{Message exchange between an FTP client (c) and lightFTP server (s) on the control channel.}
    \label{fig:10}
\end{figure}

The response code is included in the message replied to by the server. After a client sends a request to a server, the server replies with a message containing a state response code, where the state response code is used to ensure that the client's request is acknowledged and to notify the client of the current server state. The state trace of the system can be inferred by observing the state response code or some information extracted from the response. For example, the exchange of information between the FTP client (c) and the LightFTP server (s) on the control channel is described in Figure \textcolor{cyan}{\ref{fig:10}}. AFLNet \citenew{18} uses the server's response code to track the state traces of the test case, which in turn guides the fuzzer to test valid regions in the state space. Bleem \citenew{4} collects output sequences from target clients and servers at runtime. It supports and guides fuzzing by analyzing the output sequences to infer the system state and dynamically update the state space of all protocol entities (i.e., clients and servers) of the system to be tested.

\subsection{Coverage}
Coverage includes code coverage, function coverage, branch coverage, path coverage state, coverage, etc., which are generally applied to greybox fuzzers. The fuzzers perform instrumentation of the program under test and use bitmaps to record the coverage of paths, branches, etc. during test case execution. Many fuzzers such as stateAFL \citenew{stateafl}, AFLNet, and libfuzzer \citenew{74} use coverage feedback to guide fuzzing.

Anti-fuzzing techniques deceive the fuzzer by methods such as inserting pseudo-paths or adding delay and obfuscation code to the error handling code to slow down the dynamic analysis. This plays a significant preventive role against coverage-based fuzzers. In recent years, due to the widespread use of anti-fuzzing techniques, fuzzers that do not distinguish between different types of edge coverings have a harder time detecting vulnerabilities. Wang et al. \citenew{71} proposes to focus on edges associated with sensitive memory operations. It evaluates and labels edges at three levels: function, loop, and basic block, and prioritizes inputs according to new security-sensitive overlays as a way to counter anti-fuzzing techniques and detect more vulnerabilities.

\subsection{Branch Feedback and Variable Feedback}
Branch feedback is similar in principle to branch coverage in coverage, but the main difference between the two is that branch feedback requires the user to manually annotate the code. The user manually marks branches that are more likely to trigger a vulnerability as being of interest by observing them. Based on the method of monitoring branch coverage in AFL, Chen et al. \citenew{72} improves the performance of the fuzzer by performing additional observations on those branches of interest that are manually labeled. There is a category of network protocol implementations that make use of enumeration-type program variables to record information about the state of the protocol. The idea of variable feedback is precisely to obtain information about the state of the protocol by observing this type of procedural variables and to update and maintain the protocol state machine to improve the effectiveness of fuzzing. IJON \citenew{50} argues that current fuzzers are unable to properly explore the state space of programs beyond code coverage, such as those test cases that result in the same code coverage, but different state trace. By annotating those parts of the state space that should be explored further (usually with a line or two of code), programmers can help fuzzers overcome several current obstacles associated with fuzzing complex applications. The core idea is to explore the behavior of a program more systematically using program variables that represent the internal state of the program. SGFuzz \citenew{73} uses pattern matching to identify enumerated type variables as possible state variables, using variable feedback to guide fuzzing and thus explore the program state space more efficiently.

In specifically, the specific application scenarios of branch feedback and variable feedback depend on whether the PUT records the protocol state of the program as it runs via program variables or program points. In this context, program variables refer to identifiers that are used to store data in programming. Variables can contain various types of data, such as numbers, strings, booleans, etc. Program point refers to a specific location or moment used to describe the execution of a program. It may refer to a particular line in the code, the location of a function call, the location where a particular state or event occurs, etc.

\subsection{Memory Feedback}
Memory signals refer to the electrical signals or combinations of signals that are used to control the operation of and memory access (e.g., RAM, ROM, etc.) in a computer system or electronic device. They control the reading, writing, and processing of data. Some fuzzers argue that when the memory signal changes, it causes the protocol state to change as well. The core idea of memory feedback is to observe whether the current system inputs have an impact on these memory regions by taking a snapshot of the memory regions and comparing the relevant information. StateAFL \citenew{stateafl} takes snapshots of memory regions at runtime and applies a hash algorithm to map the state in each memory to a unique protocol state identifier. Through the above operations, the fuzzer can infer the current protocol state of the target server and gradually construct a protocol state machine to guide the fuzzing.

\section{Directions and Perspectives for future research}
\label{direction}
Network protocol fuzzing techniques have evolved at a high rate over the past 20 years or so, but some limitations still exist. To stimulate the design of more efficient and practical protocol fuzzers in the future, we summarise our viewpoints and thoughts on future directions.

\subsection{High Expandability}
Studies have shown (\textcolor{cyan}{\citep{11}}; \textcolor{cyan}{\citep{62}}) that any protocol object that is fuzzed for the first time reveals a large amount of new security-related findings. We believe that while focusing on improvements to the fuzzing algorithms, it is also necessary to enhance the expandability of the protocol fuzzers for the test objects. Protocol fuzzers are more dependent on syntax input than fuzzers oriented to other test objects, which also leads to the fact that existing protocol fuzzers tend to be less easy to expand with new test objects. The Peach extension requires the user to provide the full Pit file for the new protocol object, and the AFLNet extension requires the user to write C code for the new protocol response extraction. Sections \textcolor{cyan}{\ref{32}} and \textcolor{cyan}{\ref{33}} have summarised part of the methods, but these methods have their limitations and it is difficult to deal with new protocols in a convenient way. We expect fuzzing techniques to be further combined with deep learning and model inference techniques in the future.

\subsection{More Effective Feedback}
The availability of feedback information has led to the evolution of fuzzing techniques from blindness to intelligence, and AFL's use of coverage has created the world of CGF. As described in the feedback section, there are many varieties of feedback available for protocol fuzzers, but none of them do as much to bring about a milestone increase in fuzzing effectiveness as coverage. In recent years, protocol security in ICS and IoT has attracted a lot of concern, but the acquisition of coverage in test scenarios for these protocols would be limited. Is it possible to present proprietary feedback for special protocol testing scenarios (e.g., real ICS devices, IoT devices)?

\subsection{Complex Testing Scenarios}
Existing protocol fuzzers tend to consider scenarios where only one server interacts with one client. In real-world scenarios, the server can connect to a large amount of clients at the same time, i.e., one to multiple or even multiple to multiple interaction scenarios are very common, will the complex testing scenarios with multiple entities introduce new types of protocol bugs?

\subsection{Underlying Protocol Testing Capabilities}
Existing protocol fuzzers focus more on application layer protocols and only a few research efforts have been made to analyse the security of TCP and UDP. Security analysis techniques for transport layer or lower layer protocols differ from upper layer protocols and have difficulties in technical implementation. Can the function replacement and emulation techniques mentioned in Section 5 be applied to test the underlying protocols?

\section{Conclusions}
Network protocols are more challenging compared to traditional program testing due to their highly structured inputs and enormous state spaces. Fuzzing has been widely used in the field of network protocols as one of the most effective and efficient methods to discover vulnerabilities in recent years. In this paper, we summarise a unified process model for network protocol fuzzing by analyzing a large amount of high-quality literature and fuzzers and propose four concerns for network protocol fuzzing research, including protocol syntax acquiring and modeling, testcase generation, test execution and monitoring, and feedback information acquisition and utilization. Based on this model, we systematically summarise the design decisions as well as the innovations of existing protocol fuzzing techniques in the four stages of the unified process model. Finally, we propose promising directions and perspectives for future research in the area of protocol fuzzing.



\bibliographystyle{elsarticle-harv}
\bibliography{chapter/ref} 

\begin{thebibliography}{101}
\expandafter\ifx\csname natexlab\endcsname\relax\def\natexlab#1{#1}\fi
\providecommand{\url}[1]{\texttt{#1}}
\providecommand{\href}[2]{#2}
\providecommand{\path}[1]{#1}
\providecommand{\DOIprefix}{doi:}
\providecommand{\ArXivprefix}{arXiv:}
\providecommand{\URLprefix}{URL: }
\providecommand{\Pubmedprefix}{pmid:}
\providecommand{\doi}[1]{\href{http://dx.doi.org/#1}{\path{#1}}}
\providecommand{\Pubmed}[1]{\href{pmid:#1}{\path{#1}}}
\providecommand{\bibinfo}[2]{#2}
\ifx\xfnm\relax \def\xfnm[#1]{\unskip,\space#1}\fi
\bibitem[{Abdelnur et~al.(2007)Abdelnur, State and Festor}]{Kif}
\bibinfo{author}{Abdelnur, H.J.}, \bibinfo{author}{State, R.}, \bibinfo{author}{Festor, O.}, \bibinfo{year}{2007}.
\newblock \bibinfo{title}{Kif: a stateful sip fuzzer}, in: \bibinfo{booktitle}{Proceedings of the 1st international conference on Principles, systems and applications of IP telecommunications}, pp. \bibinfo{pages}{47--56}.
\bibitem[{Aitel(2002)}]{65}
\bibinfo{author}{Aitel, D.}, \bibinfo{year}{2002}.
\newblock \bibinfo{title}{The advantages of block-based protocol analysis for security testing}.
\newblock \bibinfo{journal}{Immunity Inc., February} \bibinfo{volume}{105}, \bibinfo{pages}{106}.
\bibitem[{Amini(2010)}]{64}
\bibinfo{author}{Amini}, \bibinfo{year}{2010}.
\newblock \bibinfo{title}{Sulley fuzzing framework}.
\bibitem[{Andronidis(2022)}]{14}
\bibinfo{author}{Andronidis, Anastasios, C.}, \bibinfo{year}{2022}.
\newblock \bibinfo{title}{Snapfuzz: high-throughput fuzzing of network applications}, in: \bibinfo{booktitle}{Proceedings of the 31st ACM SIGSOFT International Symposium on Software Testing and Analysis}, pp. \bibinfo{pages}{340--351}.
\bibitem[{Angluin(1987)}]{38}
\bibinfo{author}{Angluin, D.}, \bibinfo{year}{1987}.
\newblock \bibinfo{title}{Learning regular sets from queries and counterexamples}.
\newblock \bibinfo{journal}{Information and computation} \bibinfo{volume}{75}, \bibinfo{pages}{87--106}.
\bibitem[{Aschermann et~al.(2020)Aschermann, Schumilo, Abbasi and Holz}]{50}
\bibinfo{author}{Aschermann, C.}, \bibinfo{author}{Schumilo, S.}, \bibinfo{author}{Abbasi, A.}, \bibinfo{author}{Holz, T.}, \bibinfo{year}{2020}.
\newblock \bibinfo{title}{Ijon: Exploring deep state spaces via fuzzing}, in: \bibinfo{booktitle}{2020 IEEE Symposium on Security and Privacy (SP)}, \bibinfo{organization}{IEEE}. pp. \bibinfo{pages}{1597--1612}.
\bibitem[{Ba et~al.(2022)Ba, B{\"o}hme, Mirzamomen and Roychoudhury}]{73}
\bibinfo{author}{Ba, J.}, \bibinfo{author}{B{\"o}hme, M.}, \bibinfo{author}{Mirzamomen, Z.}, \bibinfo{author}{Roychoudhury, A.}, \bibinfo{year}{2022}.
\newblock \bibinfo{title}{Stateful greybox fuzzing}, in: \bibinfo{booktitle}{31st USENIX Security Symposium (USENIX Security 22)}, pp. \bibinfo{pages}{3255--3272}.
\bibitem[{Babi{\'c} et~al.(2019)Babi{\'c}, Bucur, Chen, Ivan{\v{c}}i{\'c}, King, Kusano, Lemieux, Szekeres and Wang}]{babic2019fudge}
\bibinfo{author}{Babi{\'c}, D.}, \bibinfo{author}{Bucur, S.}, \bibinfo{author}{Chen, Y.}, \bibinfo{author}{Ivan{\v{c}}i{\'c}, F.}, \bibinfo{author}{King, T.}, \bibinfo{author}{Kusano, M.}, \bibinfo{author}{Lemieux, C.}, \bibinfo{author}{Szekeres, L.}, \bibinfo{author}{Wang, W.}, \bibinfo{year}{2019}.
\newblock \bibinfo{title}{Fudge: fuzz driver generation at scale}, in: \bibinfo{booktitle}{Proceedings of the 2019 27th ACM Joint Meeting on European Software Engineering Conference and Symposium on the Foundations of Software Engineering}, pp. \bibinfo{pages}{975--985}.
\bibitem[{Banks et~al.(2006)Banks, Cova, Felmetsger, Almeroth, Kemmerer and Vigna}]{3}
\bibinfo{author}{Banks, G.}, \bibinfo{author}{Cova, M.}, \bibinfo{author}{Felmetsger, V.}, \bibinfo{author}{Almeroth, K.}, \bibinfo{author}{Kemmerer, R.}, \bibinfo{author}{Vigna, G.}, \bibinfo{year}{2006}.
\newblock \bibinfo{title}{Snooze: toward a stateful network protocol fuzzer}, in: \bibinfo{booktitle}{ISC}, \bibinfo{organization}{Springer}. pp. \bibinfo{pages}{343--358}.
\bibitem[{Beddoe(2004)}]{67}
\bibinfo{author}{Beddoe, M.A.}, \bibinfo{year}{2004}.
\newblock \bibinfo{title}{Network protocol analysis using bioinformatics algorithms}.
\newblock \bibinfo{journal}{Toorcon} \bibinfo{volume}{26}, \bibinfo{pages}{1095--1098}.
\bibitem[{B{\"o}hme and Falk(2020)}]{62}
\bibinfo{author}{B{\"o}hme, M.}, \bibinfo{author}{Falk, B.}, \bibinfo{year}{2020}.
\newblock \bibinfo{title}{Fuzzing: On the exponential cost of vulnerability discovery}, in: \bibinfo{booktitle}{Proceedings of the 28th ACM joint meeting on European software engineering conference and symposium on the foundations of software engineering}, pp. \bibinfo{pages}{713--724}.
\bibitem[{Bounimova et~al.(2013)Bounimova, Godefroid and Molnar}]{white2}
\bibinfo{author}{Bounimova, E.}, \bibinfo{author}{Godefroid, P.}, \bibinfo{author}{Molnar, D.}, \bibinfo{year}{2013}.
\newblock \bibinfo{title}{Billions and billions of constraints: Whitebox fuzz testing in production}, in: \bibinfo{booktitle}{2013 35th International Conference on Software Engineering (ICSE)}, \bibinfo{organization}{IEEE}. pp. \bibinfo{pages}{122--131}.
\bibitem[{Bratus et~al.(2008)Bratus, Hansen and Shubina}]{lzfuzz}
\bibinfo{author}{Bratus, S.}, \bibinfo{author}{Hansen, A.}, \bibinfo{author}{Shubina, A.}, \bibinfo{year}{2008}.
\newblock \bibinfo{title}{Lzfuzz: a fast compression-based fuzzer for poorly documented protocols} .
\bibitem[{Bulekov et~al.(2023)Bulekov, Das, Hajnoczi and Egele}]{bulekov2023no}
\bibinfo{author}{Bulekov, A.}, \bibinfo{author}{Das, B.}, \bibinfo{author}{Hajnoczi, S.}, \bibinfo{author}{Egele, M.}, \bibinfo{year}{2023}.
\newblock \bibinfo{title}{No grammar, no problem: Towards fuzzing the linux kernel without system-call descriptions}, in: \bibinfo{booktitle}{Network and Distributed System Security (NDSS) Symposium}.
\bibitem[{Caballero et~al.(2007)Caballero, Yin, Liang and Song}]{30}
\bibinfo{author}{Caballero, J.}, \bibinfo{author}{Yin, H.}, \bibinfo{author}{Liang, Z.}, \bibinfo{author}{Song, D.}, \bibinfo{year}{2007}.
\newblock \bibinfo{title}{Polyglot: Automatic extraction of protocol message format using dynamic binary analysis}, in: \bibinfo{booktitle}{Proceedings of the 14th ACM conference on Computer and communications security}, pp. \bibinfo{pages}{317--329}.
\bibitem[{Chen et~al.(2018a)Chen, Cui, Ma, Wu, Guo and Liu}]{chen2018systematic}
\bibinfo{author}{Chen, C.}, \bibinfo{author}{Cui, B.}, \bibinfo{author}{Ma, J.}, \bibinfo{author}{Wu, R.}, \bibinfo{author}{Guo, J.}, \bibinfo{author}{Liu, W.}, \bibinfo{year}{2018}a.
\newblock \bibinfo{title}{A systematic review of fuzzing techniques}.
\newblock \bibinfo{journal}{Computers \& Security} \bibinfo{volume}{75}, \bibinfo{pages}{118--137}.
\bibitem[{Chen et~al.(2018b)Chen, Diao, Zhao, Zuo, Lin, Wang, Lau, Sun, Yang and Zhang}]{iotfuzzer}
\bibinfo{author}{Chen, J.}, \bibinfo{author}{Diao, W.}, \bibinfo{author}{Zhao, Q.}, \bibinfo{author}{Zuo, C.}, \bibinfo{author}{Lin, Z.}, \bibinfo{author}{Wang, X.}, \bibinfo{author}{Lau, W.C.}, \bibinfo{author}{Sun, M.}, \bibinfo{author}{Yang, R.}, \bibinfo{author}{Zhang, K.}, \bibinfo{year}{2018}b.
\newblock \bibinfo{title}{Iotfuzzer: Discovering memory corruptions in iot through app-based fuzzing.}, in: \bibinfo{booktitle}{NDSS}.
\bibitem[{Chen et~al.(2019)Chen, Lan and Venkataramani}]{72}
\bibinfo{author}{Chen, Y.}, \bibinfo{author}{Lan, T.}, \bibinfo{author}{Venkataramani, G.}, \bibinfo{year}{2019}.
\newblock \bibinfo{title}{Exploring effective fuzzing strategies to analyze communication protocols}, in: \bibinfo{booktitle}{Proceedings of the 3rd ACM Workshop on Forming an Ecosystem Around Software Transformation}, pp. \bibinfo{pages}{17--23}.
\bibitem[{Comparetti et~al.(2009)Comparetti, Wondracek, Kruegel and Kirda}]{22}
\bibinfo{author}{Comparetti, P.M.}, \bibinfo{author}{Wondracek, G.}, \bibinfo{author}{Kruegel, C.}, \bibinfo{author}{Kirda, E.}, \bibinfo{year}{2009}.
\newblock \bibinfo{title}{Prospex: Protocol specification extraction}, in: \bibinfo{booktitle}{2009 30th IEEE Symposium on Security and Privacy}, \bibinfo{organization}{IEEE}. pp. \bibinfo{pages}{110--125}.
\bibitem[{Corina et~al.(2017)Corina, Machiry, Salls, Shoshitaishvili, Hao, Kruegel and Vigna}]{corina2017difuze}
\bibinfo{author}{Corina, J.}, \bibinfo{author}{Machiry, A.}, \bibinfo{author}{Salls, C.}, \bibinfo{author}{Shoshitaishvili, Y.}, \bibinfo{author}{Hao, S.}, \bibinfo{author}{Kruegel, C.}, \bibinfo{author}{Vigna, G.}, \bibinfo{year}{2017}.
\newblock \bibinfo{title}{Difuze: Interface aware fuzzing for kernel drivers}, in: \bibinfo{booktitle}{Proceedings of the 2017 ACM SIGSAC Conference on Computer and Communications Security}, pp. \bibinfo{pages}{2123--2138}.
\bibitem[{Cristy(1990)}]{20}
\bibinfo{author}{Cristy, J.}, \bibinfo{year}{1990}.
\newblock \bibinfo{title}{Imagemagick}.
\newblock \URLprefix \url{https://imagemagick.org/index.php}.
\bibitem[{Cui et~al.(2007)Cui, Kannan and Wang}]{26}
\bibinfo{author}{Cui, W.}, \bibinfo{author}{Kannan, J.}, \bibinfo{author}{Wang, H.J.}, \bibinfo{year}{2007}.
\newblock \bibinfo{title}{Discoverer: Automatic protocol reverse engineering from network traces.}, in: \bibinfo{booktitle}{USENIX Security Symposium}, pp. \bibinfo{pages}{1--14}.
\bibitem[{Cui et~al.(2006)Cui, Paxson, Weaver and Katz}]{25}
\bibinfo{author}{Cui, W.}, \bibinfo{author}{Paxson, V.}, \bibinfo{author}{Weaver, N.}, \bibinfo{author}{Katz, R.H.}, \bibinfo{year}{2006}.
\newblock \bibinfo{title}{Protocol-independent adaptive replay of application dialog.}, in: \bibinfo{booktitle}{NDSS}.
\bibitem[{Cui et~al.(2008)Cui, Peinado, Chen, Wang and Irun-Briz}]{33}
\bibinfo{author}{Cui, W.}, \bibinfo{author}{Peinado, M.}, \bibinfo{author}{Chen, K.}, \bibinfo{author}{Wang, H.J.}, \bibinfo{author}{Irun-Briz, L.}, \bibinfo{year}{2008}.
\newblock \bibinfo{title}{Tupni: Automatic reverse engineering of input formats}, in: \bibinfo{booktitle}{Proceedings of the 15th ACM conference on Computer and communications security}, pp. \bibinfo{pages}{391--402}.
\bibitem[{CVE-2014-0160(2014)}]{heartbleed}
\bibinfo{author}{CVE-2014-0160}, \bibinfo{year}{2014}.
\newblock \bibinfo{title}{Heartbleed - a vulnerability in openssl}.
\newblock \URLprefix \url{http://heartbleed.com}.
\bibitem[{Daniele et~al.(2023)Daniele, Andarzian and Poll}]{daniele2023fuzzers}
\bibinfo{author}{Daniele, C.}, \bibinfo{author}{Andarzian, S.B.}, \bibinfo{author}{Poll, E.}, \bibinfo{year}{2023}.
\newblock \bibinfo{title}{Fuzzers for stateful systems: Survey and research directions}.
\newblock \bibinfo{journal}{arXiv preprint arXiv:2301.02490} .
\bibitem[{Deng et~al.(2022)Deng, Yang, Wei and Zhang}]{freefuzz}
\bibinfo{author}{Deng, Y.}, \bibinfo{author}{Yang, C.}, \bibinfo{author}{Wei, A.}, \bibinfo{author}{Zhang, L.}, \bibinfo{year}{2022}.
\newblock \bibinfo{title}{Fuzzing deep-learning libraries via automated relational api inference}, in: \bibinfo{booktitle}{Proceedings of the 30th ACM Joint European Software Engineering Conference and Symposium on the Foundations of Software Engineering}, pp. \bibinfo{pages}{44--56}.
\bibitem[{Duchene et~al.(2014)Duchene, Rawat, Richier and Groz}]{kameleonfuzz}
\bibinfo{author}{Duchene, F.}, \bibinfo{author}{Rawat, S.}, \bibinfo{author}{Richier, J.L.}, \bibinfo{author}{Groz, R.}, \bibinfo{year}{2014}.
\newblock \bibinfo{title}{Kameleonfuzz: evolutionary fuzzing for black-box xss detection}, in: \bibinfo{booktitle}{Proceedings of the 4th ACM conference on Data and application security and privacy}, pp. \bibinfo{pages}{37--48}.
\bibitem[{Duchene et~al.(2018)Duchene, Le~Guernic, Alata, Nicomette and Ka{\^a}niche}]{23}
\bibinfo{author}{Duchene, J.}, \bibinfo{author}{Le~Guernic, C.}, \bibinfo{author}{Alata, E.}, \bibinfo{author}{Nicomette, V.}, \bibinfo{author}{Ka{\^a}niche, M.}, \bibinfo{year}{2018}.
\newblock \bibinfo{title}{State of the art of network protocol reverse engineering tools}.
\newblock \bibinfo{journal}{Journal of Computer Virology and Hacking Techniques} \bibinfo{volume}{14}, \bibinfo{pages}{53--68}.
\bibitem[{Eddington(2004)}]{15}
\bibinfo{author}{Eddington, M.}, \bibinfo{year}{2004}.
\newblock \bibinfo{title}{Peach fuzzing platform}.
\newblock \bibinfo{journal}{Peach Fuzzer} \bibinfo{volume}{34}, \bibinfo{pages}{32--43}.
\bibitem[{Fabrice~Bellard(2000)}]{19}
\bibinfo{author}{Fabrice~Bellard, B.B.}, \bibinfo{year}{2000}.
\newblock \bibinfo{title}{Ffmpeg}.
\newblock \URLprefix \url{https://ffmpeg.org/}.
\bibitem[{Feng et~al.(2021a)Feng, Sun, Zhu, Xue, Wen, Liu, Nepal and Xiang}]{snipuzz}
\bibinfo{author}{Feng, X.}, \bibinfo{author}{Sun, R.}, \bibinfo{author}{Zhu, X.}, \bibinfo{author}{Xue, M.}, \bibinfo{author}{Wen, S.}, \bibinfo{author}{Liu, D.}, \bibinfo{author}{Nepal, S.}, \bibinfo{author}{Xiang, Y.}, \bibinfo{year}{2021}a.
\newblock \bibinfo{title}{Snipuzz: Black-box fuzzing of iot firmware via message snippet inference}, in: \bibinfo{booktitle}{Proceedings of the 2021 ACM SIGSAC Conference on Computer and Communications Security}, pp. \bibinfo{pages}{337--350}.
\bibitem[{Feng et~al.(2021b)Feng, Sun, Zhu, Xue, Wen, Liu, Nepal and Xiang}]{37}
\bibinfo{author}{Feng, X.}, \bibinfo{author}{Sun, R.}, \bibinfo{author}{Zhu, X.}, \bibinfo{author}{Xue, M.}, \bibinfo{author}{Wen, S.}, \bibinfo{author}{Liu, D.}, \bibinfo{author}{Nepal, S.}, \bibinfo{author}{Xiang, Y.}, \bibinfo{year}{2021}b.
\newblock \bibinfo{title}{Snipuzz: Black-box fuzzing of iot firmware via message snippet inference}, in: \bibinfo{booktitle}{Proceedings of the 2021 ACM SIGSAC Conference on Computer and Communications Security}, pp. \bibinfo{pages}{337--350}.
\bibitem[{Ferreira et~al.(2021)Ferreira, Brewton, D'Antoni and Silva}]{42}
\bibinfo{author}{Ferreira, T.}, \bibinfo{author}{Brewton, H.}, \bibinfo{author}{D'Antoni, L.}, \bibinfo{author}{Silva, A.}, \bibinfo{year}{2021}.
\newblock \bibinfo{title}{Prognosis: closed-box analysis of network protocol implementations}, in: \bibinfo{booktitle}{Proceedings of the 2021 ACM SIGCOMM 2021 Conference}, pp. \bibinfo{pages}{762--774}.
\bibitem[{Fioraldi et~al.(2020)Fioraldi, D'Elia and Coppa}]{grey1}
\bibinfo{author}{Fioraldi, A.}, \bibinfo{author}{D'Elia, D.C.}, \bibinfo{author}{Coppa, E.}, \bibinfo{year}{2020}.
\newblock \bibinfo{title}{Weizz: Automatic grey-box fuzzing for structured binary formats}, in: \bibinfo{booktitle}{Proceedings of the 29th ACM SIGSOFT international symposium on software testing and analysis}, pp. \bibinfo{pages}{1--13}.
\bibitem[{Fiterau et~al.(2020)Fiterau, Jonsson, Merget, De~Ruiter, Sagonas and Somorovsky}]{44}
\bibinfo{author}{Fiterau, P.}, \bibinfo{author}{Jonsson, B.}, \bibinfo{author}{Merget, R.}, \bibinfo{author}{De~Ruiter, J.}, \bibinfo{author}{Sagonas, K.}, \bibinfo{author}{Somorovsky, J.}, \bibinfo{year}{2020}.
\newblock \bibinfo{title}{Analysis of dtls implementations using protocol state fuzzing}, in: \bibinfo{booktitle}{29th USENIX Security Symposium, Online, August 12--14, 2020}, pp. \bibinfo{pages}{2523--2540}.
\bibitem[{Fiter{\u{a}}u-Bro{\c{s}}tcan et~al.(2022)Fiter{\u{a}}u-Bro{\c{s}}tcan, Jonsson, Sagonas and T{\aa}quist}]{17}
\bibinfo{author}{Fiter{\u{a}}u-Bro{\c{s}}tcan, P.}, \bibinfo{author}{Jonsson, B.}, \bibinfo{author}{Sagonas, K.}, \bibinfo{author}{T{\aa}quist, F.}, \bibinfo{year}{2022}.
\newblock \bibinfo{title}{Dtls-fuzzer: A dtls protocol state fuzzer}, in: \bibinfo{booktitle}{2022 IEEE Conference on Software Testing, Verification and Validation (ICST)}, \bibinfo{organization}{IEEE}. pp. \bibinfo{pages}{456--458}.
\bibitem[{Fiter{\u{a}}u-Bro{\c{s}}tean et~al.(2016)Fiter{\u{a}}u-Bro{\c{s}}tean, Janssen and Vaandrager}]{43}
\bibinfo{author}{Fiter{\u{a}}u-Bro{\c{s}}tean, P.}, \bibinfo{author}{Janssen, R.}, \bibinfo{author}{Vaandrager, F.}, \bibinfo{year}{2016}.
\newblock \bibinfo{title}{Combining model learning and model checking to analyze tcp implementations}, in: \bibinfo{booktitle}{Computer Aided Verification: 28th International Conference, CAV 2016, Toronto, ON, Canada, July 17-23, 2016, Proceedings, Part II 28}, \bibinfo{organization}{Springer}. pp. \bibinfo{pages}{454--471}.
\bibitem[{Gascon et~al.(2015)Gascon, Wressnegger, Yamaguchi, Arp and Rieck}]{45}
\bibinfo{author}{Gascon, H.}, \bibinfo{author}{Wressnegger, C.}, \bibinfo{author}{Yamaguchi, F.}, \bibinfo{author}{Arp, D.}, \bibinfo{author}{Rieck, K.}, \bibinfo{year}{2015}.
\newblock \bibinfo{title}{Pulsar: Stateful black-box fuzzing of proprietary network protocols}, in: \bibinfo{booktitle}{Security and Privacy in Communication Networks: 11th EAI International Conference, SecureComm 2015, Dallas, TX, USA, October 26-29, 2015, Proceedings 11}, \bibinfo{organization}{Springer}. pp. \bibinfo{pages}{330--347}.
\bibitem[{Giuffrida and van~der Kouwe(2022)}]{57}
\bibinfo{author}{Giuffrida, E.G.C.}, \bibinfo{author}{van~der Kouwe, H.B.E.}, \bibinfo{year}{2022}.
\newblock \bibinfo{title}{Snappy: Efficient fuzzing with adaptive and mutable snapshots} .
\bibitem[{Godefroid(2007)}]{godefroid2007random}
\bibinfo{author}{Godefroid, P.}, \bibinfo{year}{2007}.
\newblock \bibinfo{title}{Random testing for security: blackbox vs. whitebox fuzzing}, in: \bibinfo{booktitle}{Proceedings of the 2nd international workshop on Random testing: co-located with the 22nd IEEE/ACM International Conference on Automated Software Engineering (ASE 2007)}, pp. \bibinfo{pages}{1--1}.
\bibitem[{Godefroid et~al.(2008)Godefroid, Kiezun and Levin}]{white1}
\bibinfo{author}{Godefroid, P.}, \bibinfo{author}{Kiezun, A.}, \bibinfo{author}{Levin, M.Y.}, \bibinfo{year}{2008}.
\newblock \bibinfo{title}{Grammar-based whitebox fuzzing}, in: \bibinfo{booktitle}{Proceedings of the 29th ACM SIGPLAN conference on programming language design and implementation}, pp. \bibinfo{pages}{206--215}.
\bibitem[{Gorbunov(2010)}]{46}
\bibinfo{author}{Gorbunov, S.}, \bibinfo{year}{2010}.
\newblock \bibinfo{title}{Autofuzz: Automated network protocol fuzzing framework}.
\newblock \bibinfo{journal}{Ijcsns} \bibinfo{volume}{10}, \bibinfo{pages}{239}.
\bibitem[{Hermann et~al.(1995)Hermann, Johnson, Engel and AG}]{hermann1995framework}
\bibinfo{author}{Hermann, H.}, \bibinfo{author}{Johnson, R.}, \bibinfo{author}{Engel, R.}, \bibinfo{author}{AG, A.T.}, \bibinfo{year}{1995}.
\newblock \bibinfo{title}{A framework for network protocol software}, in: \bibinfo{booktitle}{Proceedings OOPSLA ‘95, ACM SIGPLAN Notices}.
\bibitem[{Hess et~al.(1992)Hess, Safford and Pooch}]{hess1992unix}
\bibinfo{author}{Hess, D.K.}, \bibinfo{author}{Safford, D.R.}, \bibinfo{author}{Pooch, U.W.}, \bibinfo{year}{1992}.
\newblock \bibinfo{title}{A unix network protocol security study: Network information service}.
\newblock \bibinfo{journal}{ACM SIGCOMM Computer Communication Review} \bibinfo{volume}{22}, \bibinfo{pages}{24--28}.
\bibitem[{Hsu et~al.(2008)Hsu, Shu and Lee}]{47}
\bibinfo{author}{Hsu, Y.}, \bibinfo{author}{Shu, G.}, \bibinfo{author}{Lee, D.}, \bibinfo{year}{2008}.
\newblock \bibinfo{title}{A model-based approach to security flaw detection of network protocol implementations}, in: \bibinfo{booktitle}{2008 IEEE International Conference on Network Protocols}, \bibinfo{organization}{IEEE}. pp. \bibinfo{pages}{114--123}.
\bibitem[{Hu et~al.(2018)Hu, Shi, Huang, Xiong and Bu}]{35}
\bibinfo{author}{Hu, Z.}, \bibinfo{author}{Shi, J.}, \bibinfo{author}{Huang, Y.}, \bibinfo{author}{Xiong, J.}, \bibinfo{author}{Bu, X.}, \bibinfo{year}{2018}.
\newblock \bibinfo{title}{Ganfuzz: a gan-based industrial network protocol fuzzing framework}, in: \bibinfo{booktitle}{Proceedings of the 15th ACM International Conference on Computing Frontiers}, pp. \bibinfo{pages}{138--145}.
\bibitem[{Jero et~al.(2019)Jero, Pacheco, Goldwasser and Nita-Rotaru}]{34}
\bibinfo{author}{Jero, S.}, \bibinfo{author}{Pacheco, M.L.}, \bibinfo{author}{Goldwasser, D.}, \bibinfo{author}{Nita-Rotaru, C.}, \bibinfo{year}{2019}.
\newblock \bibinfo{title}{Leveraging textual specifications for grammar-based fuzzing of network protocols}, in: \bibinfo{booktitle}{Proceedings of the AAAI Conference on Artificial Intelligence}, pp. \bibinfo{pages}{9478--9483}.
\bibitem[{Kaksonen et~al.(2001)Kaksonen, Laakso and Takanen}]{63}
\bibinfo{author}{Kaksonen, R.}, \bibinfo{author}{Laakso, M.}, \bibinfo{author}{Takanen, A.}, \bibinfo{year}{2001}.
\newblock \bibinfo{title}{Software security assessment through specification mutations and fault injection}, in: \bibinfo{booktitle}{Communications and Multimedia Security Issues of the New Century: IFIP TC6/TC11 Fifth Joint Working Conference on Communications and Multimedia Security (CMS’01) May 21--22, 2001, Darmstadt, Germany}, \bibinfo{organization}{Springer}. pp. \bibinfo{pages}{173--183}.
\bibitem[{Kim et~al.(2020)Kim, Jeong, Kim, Jang, Shin and Lee}]{kim2020hfl}
\bibinfo{author}{Kim, K.}, \bibinfo{author}{Jeong, D.R.}, \bibinfo{author}{Kim, C.H.}, \bibinfo{author}{Jang, Y.}, \bibinfo{author}{Shin, I.}, \bibinfo{author}{Lee, B.}, \bibinfo{year}{2020}.
\newblock \bibinfo{title}{Hfl: Hybrid fuzzing on the linux kernel.}, in: \bibinfo{booktitle}{NDSS}.
\bibitem[{Kitagawa et~al.(2010)Kitagawa, Hanaoka and Kono}]{aspfuzz}
\bibinfo{author}{Kitagawa, T.}, \bibinfo{author}{Hanaoka, M.}, \bibinfo{author}{Kono, K.}, \bibinfo{year}{2010}.
\newblock \bibinfo{title}{Aspfuzz: A state-aware protocol fuzzer based on application-layer protocols}, in: \bibinfo{booktitle}{The IEEE symposium on Computers and Communications}, \bibinfo{organization}{IEEE}. pp. \bibinfo{pages}{202--208}.
\bibitem[{Li et~al.(2022)Li, Li, Sun, Chen and Yu}]{13}
\bibinfo{author}{Li, J.}, \bibinfo{author}{Li, S.}, \bibinfo{author}{Sun, G.}, \bibinfo{author}{Chen, T.}, \bibinfo{author}{Yu, H.}, \bibinfo{year}{2022}.
\newblock \bibinfo{title}{Snpsfuzzer: A fast greybox fuzzer for stateful network protocols using snapshots}.
\newblock \bibinfo{journal}{IEEE Transactions on Information Forensics and Security} \bibinfo{volume}{17}, \bibinfo{pages}{2673--2687}.
\bibitem[{Li et~al.(2018)Li, Zhao and Zhang}]{li2018fuzzing}
\bibinfo{author}{Li, J.}, \bibinfo{author}{Zhao, B.}, \bibinfo{author}{Zhang, C.}, \bibinfo{year}{2018}.
\newblock \bibinfo{title}{Fuzzing: a survey}.
\newblock \bibinfo{journal}{Cybersecurity} \bibinfo{volume}{1}, \bibinfo{pages}{1--13}.
\bibitem[{Li et~al.(2021)Li, Li, Fu, Xue, Yu and Sun}]{75}
\bibinfo{author}{Li, S.}, \bibinfo{author}{Li, J.}, \bibinfo{author}{Fu, J.}, \bibinfo{author}{Xue, M.}, \bibinfo{author}{Yu, H.}, \bibinfo{author}{Sun, G.}, \bibinfo{year}{2021}.
\newblock \bibinfo{title}{Protocol fuzzing with specification guided message generation}, in: \bibinfo{booktitle}{2021 International Conference on UK-China Emerging Technologies (UCET)}, \bibinfo{organization}{IEEE}. pp. \bibinfo{pages}{164--170}.
\bibitem[{Liang et~al.(2018)Liang, Pei, Jia, Shen and Zhang}]{59}
\bibinfo{author}{Liang, H.}, \bibinfo{author}{Pei, X.}, \bibinfo{author}{Jia, X.}, \bibinfo{author}{Shen, W.}, \bibinfo{author}{Zhang, J.}, \bibinfo{year}{2018}.
\newblock \bibinfo{title}{Fuzzing: State of the art}.
\newblock \bibinfo{journal}{IEEE Transactions on Reliability} \bibinfo{volume}{67}, \bibinfo{pages}{1199--1218}.
\bibitem[{Lin et~al.(2008)Lin, Jiang, Xu and Zhang}]{31}
\bibinfo{author}{Lin, Z.}, \bibinfo{author}{Jiang, X.}, \bibinfo{author}{Xu, D.}, \bibinfo{author}{Zhang, X.}, \bibinfo{year}{2008}.
\newblock \bibinfo{title}{Automatic protocol format reverse engineering through context-aware monitored execution.}, in: \bibinfo{booktitle}{NDSS}, pp. \bibinfo{pages}{1--15}.
\bibitem[{Liu et~al.(2023)Liu, Toffalini, Zhou and Payer}]{liu2023videzzo}
\bibinfo{author}{Liu, Q.}, \bibinfo{author}{Toffalini, F.}, \bibinfo{author}{Zhou, Y.}, \bibinfo{author}{Payer, M.}, \bibinfo{year}{2023}.
\newblock \bibinfo{title}{Videzzo: Dependency-aware virtual device fuzzing}, in: \bibinfo{booktitle}{2023 IEEE Symposium on Security and Privacy (SP)}, \bibinfo{organization}{IEEE Computer Society}. pp. \bibinfo{pages}{3228--3245}.
\bibitem[{Luo et~al.(2023)Luo, Yu, Zuo, Liu, Jiang, Chen, Roychoudhury and Sun}]{4}
\bibinfo{author}{Luo, Z.}, \bibinfo{author}{Yu, J.}, \bibinfo{author}{Zuo, F.}, \bibinfo{author}{Liu, J.}, \bibinfo{author}{Jiang, Y.}, \bibinfo{author}{Chen, T.}, \bibinfo{author}{Roychoudhury, A.}, \bibinfo{author}{Sun, J.}, \bibinfo{year}{2023}.
\newblock \bibinfo{title}{Bleem: Packet sequence oriented fuzzing for protocol implementations} .
\bibitem[{Luo et~al.(2019)Luo, Zuo, Jiang, Gao, Jiao and Sun}]{polar}
\bibinfo{author}{Luo, Z.}, \bibinfo{author}{Zuo, F.}, \bibinfo{author}{Jiang, Y.}, \bibinfo{author}{Gao, J.}, \bibinfo{author}{Jiao, X.}, \bibinfo{author}{Sun, J.}, \bibinfo{year}{2019}.
\newblock \bibinfo{title}{Polar: Function code aware fuzz testing of ics protocol}.
\newblock \bibinfo{journal}{ACM Transactions on Embedded Computing Systems (TECS)} \bibinfo{volume}{18}, \bibinfo{pages}{1--22}.
\bibitem[{Luo et~al.(2020)Luo, Zuo, Shen, Jiao, Chang and Jiang}]{66}
\bibinfo{author}{Luo, Z.}, \bibinfo{author}{Zuo, F.}, \bibinfo{author}{Shen, Y.}, \bibinfo{author}{Jiao, X.}, \bibinfo{author}{Chang, W.}, \bibinfo{author}{Jiang, Y.}, \bibinfo{year}{2020}.
\newblock \bibinfo{title}{Ics protocol fuzzing: coverage guided packet crack and generation}, in: \bibinfo{booktitle}{2020 57th ACM/IEEE Design Automation Conference (DAC)}, \bibinfo{organization}{IEEE}. pp. \bibinfo{pages}{1--6}.
\bibitem[{Maier et~al.(2022)Maier, Bittner, Munier and Beier}]{69}
\bibinfo{author}{Maier, D.}, \bibinfo{author}{Bittner, O.}, \bibinfo{author}{Munier, M.}, \bibinfo{author}{Beier, J.}, \bibinfo{year}{2022}.
\newblock \bibinfo{title}{Fitm: Binary-only coverage-guided fuzzing for stateful network protocols}, in: \bibinfo{booktitle}{Workshop on Binary Analysis Research (BAR)}.
\bibitem[{Man{\`e}s et~al.(2019)Man{\`e}s, Han, Han, Cha, Egele, Schwartz and Woo}]{58}
\bibinfo{author}{Man{\`e}s, V.J.}, \bibinfo{author}{Han, H.}, \bibinfo{author}{Han, C.}, \bibinfo{author}{Cha, S.K.}, \bibinfo{author}{Egele, M.}, \bibinfo{author}{Schwartz, E.J.}, \bibinfo{author}{Woo, M.}, \bibinfo{year}{2019}.
\newblock \bibinfo{title}{The art, science, and engineering of fuzzing: A survey}.
\newblock \bibinfo{journal}{IEEE Transactions on Software Engineering} \bibinfo{volume}{47}, \bibinfo{pages}{2312--2331}.
\bibitem[{Miller et~al.(1990)Miller, Fredriksen and So}]{8}
\bibinfo{author}{Miller, B.P.}, \bibinfo{author}{Fredriksen, L.}, \bibinfo{author}{So, B.}, \bibinfo{year}{1990}.
\newblock \bibinfo{title}{An empirical study of the reliability of unix utilities}.
\newblock \bibinfo{journal}{Communications of the ACM} \bibinfo{volume}{33}, \bibinfo{pages}{32--44}.
\bibitem[{Munea et~al.(2016)Munea, Lim and Shon}]{munea2016network}
\bibinfo{author}{Munea, T.L.}, \bibinfo{author}{Lim, H.}, \bibinfo{author}{Shon, T.}, \bibinfo{year}{2016}.
\newblock \bibinfo{title}{Network protocol fuzz testing for information systems and applications: a survey and taxonomy}.
\newblock \bibinfo{journal}{Multimedia tools and applications} \bibinfo{volume}{75}, \bibinfo{pages}{14745--14757}.
\bibitem[{Narayan et~al.(2015)Narayan, Shukla and Clancy}]{24}
\bibinfo{author}{Narayan, J.}, \bibinfo{author}{Shukla, S.K.}, \bibinfo{author}{Clancy, T.C.}, \bibinfo{year}{2015}.
\newblock \bibinfo{title}{A survey of automatic protocol reverse engineering tools}.
\newblock \bibinfo{journal}{ACM Computing Surveys (CSUR)} \bibinfo{volume}{48}, \bibinfo{pages}{1--26}.
\bibitem[{Natella(2022)}]{stateafl}
\bibinfo{author}{Natella, R.}, \bibinfo{year}{2022}.
\newblock \bibinfo{title}{Stateafl: Greybox fuzzing for stateful network servers}.
\newblock \bibinfo{journal}{Empirical Software Engineering} \bibinfo{volume}{27}, \bibinfo{pages}{191}.
\bibitem[{Oehlert(2005)}]{6}
\bibinfo{author}{Oehlert, P.}, \bibinfo{year}{2005}.
\newblock \bibinfo{title}{Violating assumptions with fuzzing}.
\newblock \bibinfo{journal}{IEEE Security \& Privacy} \bibinfo{volume}{3}, \bibinfo{pages}{58--62}.
\bibitem[{Pan et~al.(2021)Pan, Lin, Zhang, Jia, Ji, Wu, Ying, Wang and Wu}]{pan2021v}
\bibinfo{author}{Pan, G.}, \bibinfo{author}{Lin, X.}, \bibinfo{author}{Zhang, X.}, \bibinfo{author}{Jia, Y.}, \bibinfo{author}{Ji, S.}, \bibinfo{author}{Wu, C.}, \bibinfo{author}{Ying, X.}, \bibinfo{author}{Wang, J.}, \bibinfo{author}{Wu, Y.}, \bibinfo{year}{2021}.
\newblock \bibinfo{title}{V-shuttle: Scalable and semantics-aware hypervisor virtual device fuzzing}, in: \bibinfo{booktitle}{Proceedings of the 2021 ACM SIGSAC Conference on Computer and Communications Security}, pp. \bibinfo{pages}{2197--2213}.
\bibitem[{Pereyda(2015)}]{boofuzz}
\bibinfo{author}{Pereyda, J.}, \bibinfo{year}{2015}.
\newblock \bibinfo{title}{boofuzz: A fork and successor of the sulley fuzzing framework}.
\bibitem[{Pham et~al.(2020)Pham, B{\"o}hme and Roychoudhury}]{18}
\bibinfo{author}{Pham, V.T.}, \bibinfo{author}{B{\"o}hme, M.}, \bibinfo{author}{Roychoudhury, A.}, \bibinfo{year}{2020}.
\newblock \bibinfo{title}{Aflnet: a greybox fuzzer for network protocols}, in: \bibinfo{booktitle}{2020 IEEE 13th International Conference on Software Testing, Validation and Verification (ICST)}, \bibinfo{organization}{IEEE}. pp. \bibinfo{pages}{460--465}.
\bibitem[{Postel and Reynolds(1997)}]{21}
\bibinfo{author}{Postel, J.}, \bibinfo{author}{Reynolds, J.}, \bibinfo{year}{1997}.
\newblock \bibinfo{title}{Instructions to RFC authors}.
\newblock \bibinfo{type}{Technical Report}.
\bibitem[{Qin et~al.(2023)Qin, Hu, Ma, Zhao, Yin and Zhang}]{81}
\bibinfo{author}{Qin, S.}, \bibinfo{author}{Hu, F.}, \bibinfo{author}{Ma, Z.}, \bibinfo{author}{Zhao, B.}, \bibinfo{author}{Yin, T.}, \bibinfo{author}{Zhang, C.}, \bibinfo{year}{2023}.
\newblock \bibinfo{title}{Nsfuzz: Towards efficient and state-aware network service fuzzing}.
\newblock \bibinfo{journal}{ACM Transactions on Software Engineering and Methodology} .
\bibitem[{Raffelt et~al.(2009a)Raffelt, Merten, Steffen and Margaria}]{48}
\bibinfo{author}{Raffelt, H.}, \bibinfo{author}{Merten, M.}, \bibinfo{author}{Steffen, B.}, \bibinfo{author}{Margaria, T.}, \bibinfo{year}{2009}a.
\newblock \bibinfo{title}{Dynamic testing via automata learning}.
\newblock \bibinfo{journal}{International journal on software tools for technology transfer} \bibinfo{volume}{11}, \bibinfo{pages}{307--324}.
\bibitem[{Raffelt et~al.(2009b)Raffelt, Steffen, Berg and Margaria}]{39}
\bibinfo{author}{Raffelt, H.}, \bibinfo{author}{Steffen, B.}, \bibinfo{author}{Berg, T.}, \bibinfo{author}{Margaria, T.}, \bibinfo{year}{2009}b.
\newblock \bibinfo{title}{Learnlib: a framework for extrapolating behavioral models}.
\newblock \bibinfo{journal}{International journal on software tools for technology transfer} \bibinfo{volume}{11}, \bibinfo{pages}{393--407}.
\bibitem[{Redini et~al.(2021)Redini, Continella, Das, De~Pasquale, Spahn, Machiry, Bianchi, Kruegel and Vigna}]{diane}
\bibinfo{author}{Redini, N.}, \bibinfo{author}{Continella, A.}, \bibinfo{author}{Das, D.}, \bibinfo{author}{De~Pasquale, G.}, \bibinfo{author}{Spahn, N.}, \bibinfo{author}{Machiry, A.}, \bibinfo{author}{Bianchi, A.}, \bibinfo{author}{Kruegel, C.}, \bibinfo{author}{Vigna, G.}, \bibinfo{year}{2021}.
\newblock \bibinfo{title}{Diane: Identifying fuzzing triggers in apps to generate under-constrained inputs for iot devices}, in: \bibinfo{booktitle}{2021 IEEE Symposium on Security and Privacy (SP)}, \bibinfo{organization}{IEEE}. pp. \bibinfo{pages}{484--500}.
\bibitem[{Ruiter and Joeri(2015)}]{40}
\bibinfo{author}{Ruiter, D.}, \bibinfo{author}{Joeri}, \bibinfo{year}{2015}.
\newblock \bibinfo{title}{Protocol state fuzzing of $\{$TLS$\}$ implementations}, in: \bibinfo{booktitle}{24th $\{$USENIX$\}$ Security Symposium ($\{$USENIX$\}$ Security 15)}, pp. \bibinfo{pages}{193--206}.
\bibitem[{Schumilo et~al.(2021)Schumilo, Aschermann, Abbasi, W{\"o}rner and Holz}]{53}
\bibinfo{author}{Schumilo, S.}, \bibinfo{author}{Aschermann, C.}, \bibinfo{author}{Abbasi, A.}, \bibinfo{author}{W{\"o}rner, S.}, \bibinfo{author}{Holz, T.}, \bibinfo{year}{2021}.
\newblock \bibinfo{title}{Nyx: Greybox hypervisor fuzzing using fast snapshots and affine types.}, in: \bibinfo{booktitle}{USENIX Security Symposium}, pp. \bibinfo{pages}{2597--2614}.
\bibitem[{Schumilo et~al.(2017)Schumilo, Aschermann, Gawlik, Schinzel and Holz}]{54}
\bibinfo{author}{Schumilo, S.}, \bibinfo{author}{Aschermann, C.}, \bibinfo{author}{Gawlik, R.}, \bibinfo{author}{Schinzel, S.}, \bibinfo{author}{Holz, T.}, \bibinfo{year}{2017}.
\newblock \bibinfo{title}{kafl: Hardware-assisted feedback fuzzing for os kernels.}, in: \bibinfo{booktitle}{USENIX Security Symposium}, pp. \bibinfo{pages}{167--182}.
\bibitem[{Schumilo et~al.(2022)Schumilo, Aschermann, Jemmett, Abbasi and Holz}]{11}
\bibinfo{author}{Schumilo, S.}, \bibinfo{author}{Aschermann, C.}, \bibinfo{author}{Jemmett, A.}, \bibinfo{author}{Abbasi, A.}, \bibinfo{author}{Holz, T.}, \bibinfo{year}{2022}.
\newblock \bibinfo{title}{Nyx-net: network fuzzing with incremental snapshots}, in: \bibinfo{booktitle}{Proceedings of the Seventeenth European Conference on Computer Systems}, pp. \bibinfo{pages}{166--180}.
\bibitem[{Serebryany(2015)}]{74}
\bibinfo{author}{Serebryany, K.}, \bibinfo{year}{2015}.
\newblock \bibinfo{title}{libfuzzer--a library for coverage-guided fuzz testing}.
\newblock \bibinfo{journal}{LLVM project} .
\bibitem[{Serebryany(2017)}]{5}
\bibinfo{author}{Serebryany, K.}, \bibinfo{year}{2017}.
\newblock \bibinfo{title}{Oss-fuzz-google’s continuous fuzzing service for open source software}, in: \bibinfo{booktitle}{USENIX Security symposium}, \bibinfo{organization}{USENIX Association}.
\bibitem[{Shi et~al.(2023)Shi, Wang, Feng, Lan, Qin, You, Zou, Payer and Zhang}]{aifore}
\bibinfo{author}{Shi, J.}, \bibinfo{author}{Wang, Z.}, \bibinfo{author}{Feng, Z.}, \bibinfo{author}{Lan, Y.}, \bibinfo{author}{Qin, S.}, \bibinfo{author}{You, W.}, \bibinfo{author}{Zou, W.}, \bibinfo{author}{Payer, M.}, \bibinfo{author}{Zhang, C.}, \bibinfo{year}{2023}.
\newblock \bibinfo{title}{$\{$AIFORE$\}$: Smart fuzzing based on automatic input format reverse engineering}, in: \bibinfo{booktitle}{32nd USENIX Security Symposium (USENIX Security 23)}, pp. \bibinfo{pages}{4967--4984}.
\bibitem[{Sklenar(2011)}]{proxyfuzz}
\bibinfo{author}{Sklenar, P.}, \bibinfo{year}{2011}.
\newblock \bibinfo{title}{The proxyfuzz project}.
\newblock \URLprefix \url{https://src.fedoraproject.org/rpms/proxyfuzz}.
\bibitem[{Somorovsky(2016a)}]{16}
\bibinfo{author}{Somorovsky, J.}, \bibinfo{year}{2016}a.
\newblock \bibinfo{title}{Systematic fuzzing and testing of tls libraries}, in: \bibinfo{booktitle}{Proceedings of the 2016 ACM SIGSAC conference on computer and communications security}, pp. \bibinfo{pages}{1492--1504}.
\bibitem[{Somorovsky(2016b)}]{41}
\bibinfo{author}{Somorovsky, J.}, \bibinfo{year}{2016}b.
\newblock \bibinfo{title}{Systematic fuzzing and testing of tls libraries}, in: \bibinfo{booktitle}{Proceedings of the 2016 ACM SIGSAC conference on computer and communications security}, pp. \bibinfo{pages}{1492--1504}.
\bibitem[{Song et~al.(2020)Song, Hetzelt, Kim, Kang, Seifert and Franz}]{55}
\bibinfo{author}{Song, D.}, \bibinfo{author}{Hetzelt, F.}, \bibinfo{author}{Kim, J.}, \bibinfo{author}{Kang, B.B.}, \bibinfo{author}{Seifert, J.P.}, \bibinfo{author}{Franz, M.}, \bibinfo{year}{2020}.
\newblock \bibinfo{title}{Agamotto: Accelerating kernel driver fuzzing with lightweight virtual machine checkpoints}, in: \bibinfo{booktitle}{Proceedings of the 29th USENIX Conference on Security Symposium}, pp. \bibinfo{pages}{2541--2557}.
\bibitem[{Sun et~al.(2022)Sun, Lv, You, Sun, Chen, Zheng and Sun}]{27}
\bibinfo{author}{Sun, Y.}, \bibinfo{author}{Lv, S.}, \bibinfo{author}{You, J.}, \bibinfo{author}{Sun, Y.}, \bibinfo{author}{Chen, X.}, \bibinfo{author}{Zheng, Y.}, \bibinfo{author}{Sun, L.}, \bibinfo{year}{2022}.
\newblock \bibinfo{title}{Ipspex: Enabling efficient fuzzing via specification extraction on ics protocol}, in: \bibinfo{booktitle}{Applied Cryptography and Network Security: 20th International Conference, ACNS 2022, Rome, Italy, June 20--23, 2022, Proceedings}, \bibinfo{organization}{Springer}. pp. \bibinfo{pages}{356--375}.
\bibitem[{Tsankov et~al.(2012)Tsankov, Dashti and Basin}]{70}
\bibinfo{author}{Tsankov, P.}, \bibinfo{author}{Dashti, M.T.}, \bibinfo{author}{Basin, D.}, \bibinfo{year}{2012}.
\newblock \bibinfo{title}{Secfuzz: Fuzz-testing security protocols}, in: \bibinfo{booktitle}{2012 7th International Workshop on Automation of Software Test (AST)}, \bibinfo{organization}{IEEE}. pp. \bibinfo{pages}{1--7}.
\bibitem[{Wang et~al.(2019)Wang, Chen, Wei and Liu}]{grey2}
\bibinfo{author}{Wang, J.}, \bibinfo{author}{Chen, B.}, \bibinfo{author}{Wei, L.}, \bibinfo{author}{Liu, Y.}, \bibinfo{year}{2019}.
\newblock \bibinfo{title}{Superion: Grammar-aware greybox fuzzing}, in: \bibinfo{booktitle}{2019 IEEE/ACM 41st International Conference on Software Engineering (ICSE)}, \bibinfo{organization}{IEEE}. pp. \bibinfo{pages}{724--735}.
\bibitem[{Wang et~al.(2020)Wang, Jia, Liu, Zeng, Bao, Wu and Su}]{71}
\bibinfo{author}{Wang, Y.}, \bibinfo{author}{Jia, X.}, \bibinfo{author}{Liu, Y.}, \bibinfo{author}{Zeng, K.}, \bibinfo{author}{Bao, T.}, \bibinfo{author}{Wu, D.}, \bibinfo{author}{Su, P.}, \bibinfo{year}{2020}.
\newblock \bibinfo{title}{Not all coverage measurements are equal: Fuzzing by coverage accounting for input prioritization.}, in: \bibinfo{booktitle}{NDSS}.
\bibitem[{Wei et~al.(2022)Wei, Deng, Yang and Zhang}]{wei2022free}
\bibinfo{author}{Wei, A.}, \bibinfo{author}{Deng, Y.}, \bibinfo{author}{Yang, C.}, \bibinfo{author}{Zhang, L.}, \bibinfo{year}{2022}.
\newblock \bibinfo{title}{Free lunch for testing: Fuzzing deep-learning libraries from open source}, in: \bibinfo{booktitle}{Proceedings of the 44th International Conference on Software Engineering}, pp. \bibinfo{pages}{995--1007}.
\bibitem[{Wondracek et~al.(2008)Wondracek, Comparetti, Kruegel, Kirda and Anna}]{32}
\bibinfo{author}{Wondracek, G.}, \bibinfo{author}{Comparetti, P.M.}, \bibinfo{author}{Kruegel, C.}, \bibinfo{author}{Kirda, E.}, \bibinfo{author}{Anna, S.S.S.}, \bibinfo{year}{2008}.
\newblock \bibinfo{title}{Automatic network protocol analysis.}, in: \bibinfo{booktitle}{NDSS}, \bibinfo{organization}{Citeseer}. pp. \bibinfo{pages}{1--14}.
\bibitem[{Xu et~al.(2017)Xu, Kashyap, Min and Kim}]{52}
\bibinfo{author}{Xu, W.}, \bibinfo{author}{Kashyap, S.}, \bibinfo{author}{Min, C.}, \bibinfo{author}{Kim, T.}, \bibinfo{year}{2017}.
\newblock \bibinfo{title}{Designing new operating primitives to improve fuzzing performance}, in: \bibinfo{booktitle}{Proceedings of the 2017 ACM SIGSAC Conference on Computer and Communications Security}, pp. \bibinfo{pages}{2313--2328}.
\bibitem[{You et~al.(2019)You, Wang, Ma, Huang, Zhang, Wang and Liang}]{profuzzer}
\bibinfo{author}{You, W.}, \bibinfo{author}{Wang, X.}, \bibinfo{author}{Ma, S.}, \bibinfo{author}{Huang, J.}, \bibinfo{author}{Zhang, X.}, \bibinfo{author}{Wang, X.}, \bibinfo{author}{Liang, B.}, \bibinfo{year}{2019}.
\newblock \bibinfo{title}{Profuzzer: On-the-fly input type probing for better zero-day vulnerability discovery}, in: \bibinfo{booktitle}{2019 IEEE symposium on security and privacy (SP)}, \bibinfo{organization}{IEEE}. pp. \bibinfo{pages}{769--786}.
\bibitem[{Yun et~al.(2022)Yun, Rustamov, Kim and Shin}]{yun2022fuzzing}
\bibinfo{author}{Yun, J.}, \bibinfo{author}{Rustamov, F.}, \bibinfo{author}{Kim, J.}, \bibinfo{author}{Shin, Y.}, \bibinfo{year}{2022}.
\newblock \bibinfo{title}{Fuzzing of embedded systems: A survey}.
\newblock \bibinfo{journal}{ACM Computing Surveys} \bibinfo{volume}{55}, \bibinfo{pages}{1--33}.
\bibitem[{Zalewski(2017)}]{12}
\bibinfo{author}{Zalewski, M.}, \bibinfo{year}{2017}.
\newblock \bibinfo{title}{American fuzzy lop}.
\bibitem[{Zhang et~al.(2021)Zhang, Lin, Li, Xue, Xie, Chen, Ying, Wang and Liu}]{zhang2021apicraft}
\bibinfo{author}{Zhang, C.}, \bibinfo{author}{Lin, X.}, \bibinfo{author}{Li, Y.}, \bibinfo{author}{Xue, Y.}, \bibinfo{author}{Xie, J.}, \bibinfo{author}{Chen, H.}, \bibinfo{author}{Ying, X.}, \bibinfo{author}{Wang, J.}, \bibinfo{author}{Liu, Y.}, \bibinfo{year}{2021}.
\newblock \bibinfo{title}{$\{$APICraft$\}$: Fuzz driver generation for closed-source $\{$SDK$\}$ libraries}, in: \bibinfo{booktitle}{30th USENIX Security Symposium (USENIX Security 21)}, pp. \bibinfo{pages}{2811--2828}.
\bibitem[{Zhao et~al.(2019)Zhao, Li, Wei, Shi and Huang}]{36}
\bibinfo{author}{Zhao, H.}, \bibinfo{author}{Li, Z.}, \bibinfo{author}{Wei, H.}, \bibinfo{author}{Shi, J.}, \bibinfo{author}{Huang, Y.}, \bibinfo{year}{2019}.
\newblock \bibinfo{title}{Seqfuzzer: An industrial protocol fuzzing framework from a deep learning perspective}, in: \bibinfo{booktitle}{2019 12th IEEE Conference on software testing, validation and verification (ICST)}, \bibinfo{organization}{IEEE}. pp. \bibinfo{pages}{59--67}.
\bibitem[{Zhu et~al.(2022)Zhu, Wen, Camtepe and Xiang}]{1}
\bibinfo{author}{Zhu, X.}, \bibinfo{author}{Wen, S.}, \bibinfo{author}{Camtepe, S.}, \bibinfo{author}{Xiang, Y.}, \bibinfo{year}{2022}.
\newblock \bibinfo{title}{Fuzzing: a survey for roadmap}.
\newblock \bibinfo{journal}{ACM Computing Surveys (CSUR)} \bibinfo{volume}{54}, \bibinfo{pages}{1--36}.
\bibitem[{Zuo et~al.(2022)Zuo, Luo, Yu, Chen, Xu, Cui and Jiang}]{56}
\bibinfo{author}{Zuo, F.}, \bibinfo{author}{Luo, Z.}, \bibinfo{author}{Yu, J.}, \bibinfo{author}{Chen, T.}, \bibinfo{author}{Xu, Z.}, \bibinfo{author}{Cui, A.}, \bibinfo{author}{Jiang, Y.}, \bibinfo{year}{2022}.
\newblock \bibinfo{title}{Vulnerability detection of ics protocols via cross-state fuzzing}.
\newblock \bibinfo{journal}{IEEE Transactions on Computer-Aided Design of Integrated Circuits and Systems} \bibinfo{volume}{41}, \bibinfo{pages}{4457--4468}.
\bibitem[{Zuo et~al.(2021)Zuo, Luo, Yu, Liu and Jiang}]{76}
\bibinfo{author}{Zuo, F.}, \bibinfo{author}{Luo, Z.}, \bibinfo{author}{Yu, J.}, \bibinfo{author}{Liu, Z.}, \bibinfo{author}{Jiang, Y.}, \bibinfo{year}{2021}.
\newblock \bibinfo{title}{Pavfuzz: State-sensitive fuzz testing of protocols in autonomous vehicles}, in: \bibinfo{booktitle}{2021 58th ACM/IEEE Design Automation Conference (DAC)}, \bibinfo{organization}{IEEE}. pp. \bibinfo{pages}{823--828}.

\end{thebibliography}

\end{document}